\documentclass[aps,pra,reprint,superscriptaddress]{revtex4-2}

\usepackage{graphicx}
\usepackage{xcolor}
\usepackage{amsmath,esint}
\usepackage{amssymb,bm}
\usepackage{gensymb}
\usepackage{braket}
\usepackage{enumitem}
\usepackage{float}
\usepackage{tikz}
\usepackage[colorlinks=true,linkcolor=blue,citecolor=red]{hyperref}
\usepackage{mathtools}
\usepackage{tcolorbox}
\usepackage{relsize}
\usepackage{multirow}

\newcommand{\Li}{\mathcal{L}}
\newcommand{\js}{\widetilde{\sigma}}
\newcommand{\jc}{\widetilde{\xi}}
\newcommand{\Cgc}{\mathcal{C}_{\textsc{gc}}}
\newcommand{\Cnc}{\mathcal{C}_{\textsc{nc}}}
\newcommand{\Qc}{\mathcal{Q}_{\textsc{c}}}
\newcommand{\Cg}{\mathcal{C}_{\textsc{gs}}}
\newcommand{\Cn}{\mathcal{C}_{\textsc{ns}}}
\newcommand{\Qs}{\mathcal{Q}_{\textsc{s}}}

\newcommand{\rmd}{\text{d}}

\newcommand{\Gs}{\Gamma}
\newcommand{\aop}[1]{\hat{a}_{#1}}
\newcommand{\ad}[1]{\hat{a}^{\dagger}_{#1}}
\newcommand{\aopm}[2]{\hat{a}^{#2}_{#1}}
\newcommand{\adm}[2]{\hat{a}^{\dagger #2}_{#1}}
\newcommand{\eref}[1]{(\ref{#1})}
\graphicspath{{./figures/},{./figures/old/.}}

\newcommand{\be}{\begin{equation}}
\newcommand{\ee}{\end{equation}}
\newcommand{\ben}{\begin{eqnarray}}
\newcommand{\een}{\end{eqnarray}}
\newcommand{\bF}{\begin{figure}}
\newcommand{\eF}{\end{figure}}
\newcommand{\bi}{\begin{itemize}}
\newcommand{\ei}{\end{itemize}}

\usepgfmodule{nonlineartransformations}
\tikzset{lasernode/.style={
    top color=red,
    bottom color=red,
    middle color = white
    }
}

\usetikzlibrary{decorations.pathmorphing}

\tikzset{snake it/.style={decorate, decoration=snake}}

\begin{document}

\title{Estimating spacetime fluctuation strength in SU(1,1) and SU(2) interferometers}

\author{B.~Sharmila}
\email{Sharmila.Balamurugan@warwick.ac.uk}
\affiliation{Department of Physics, University of Warwick, Coventry CV4 7AL, UK.}

\date{\today}

\begin{abstract}
High-precision laser interferometers are commonly used to search for signatures of random spacetime fluctuations, in an attempt to understand the fundamental nature of gravity. Conventionally, $SU(2)$ interferometers have been used in such experimental investigations. Motivated by [K. Zheng \textit{et al.}, Photon. Res. \textbf{8}, 1653 (2020)], I assess if there is any advantage to be gained by using an $SU(1,1)$ interferometer instead of the $SU(2)$ interferometer. To this end, I compute the quantum Fisher information for estimating the strength and the correlation length of the spacetime fluctuations, and the corresponding classical Fisher information considering different experimentally relevant measurement schemes, in both types of interferometers. I compare the information metrics corresponding to different parameter regimes that are relevant to both current and possible future experimental setups. This helps me assess if and when the $SU(1,1)$ interferometer offers any advantage over the $SU(2)$ interferometer for estimating either the fluctuation strength or correlation length of the spacetime fluctuations. 
\end{abstract}


\maketitle


\section{Introduction}

Phase estimation using laser interferometers has been extensively examined in the context of gravitational wave detection for over half a century (see, for instance,~\cite{Caves81,Caves85P1,Caves85P2,GravWavIntRev,EntangPhaseEst}). Quantum Fisher information (QFI)~\cite{QFIhels,QFIparis}, that provides the ultimate limit of precision, in estimating the phase difference between the interferometer arms has been computed~\cite{Caves81,QFIinterfer,QFIintOneMode26}. This limit of precision is also achieved using suitable measurement schemes~\cite{IntOptMeas} and chosen optical states at the two input ports~\cite{IntOptState}. The ultra-high precision in phase estimation offered by laser interferometers has extended their utility to even searches for signatures that help in understanding the fundamental nature of gravity.  This has lead to many proposals~\cite{amelino99,amelinoPRD} and subsequent experimental investigations~\cite{HoloData,bentHolo,QUESTdata} that use these interferometers to search for random spacetime fluctuations, a \textit{leit-motif} in many quantum and semiclassical models of gravity. Prior work on identifying a suitable interferometric setup that can capture spacetime fluctuations, considered the estimation of a constant phase \cite{QLtIRB1,QLtIRB2}. However, contemporary work~\cite{gardner2025} has indicated the need to model stochastic fluctuations in phase and to estimate the standard deviation of these fluctuations. Ref.~\cite{gardner2025} considers a simple Gaussian noise channel to model stochastic fluctuations in a single optical mode corresponding to one of the output ports of the interferometer. However, modelling both output ports of the conventional interferometer could reveal crucial conditions that relate the system parameters with the characteristics of the random spacetime fluctuations. Further, this would be convenient when I attempt to compare the QFI obtained in the context of the conventional interferometer with that obtained in the context of any modified interferometric setup.

The enhancement in precision provided by squeezed input light in conventional interferometers naturally led to the idea of $SU(1,1)$ interferometry~\cite{SU11YMK,SU11Caves}. A conventional interferometer, characterized by an $SU(2)$ transformation of the electromagnetic field, has a half-silvered mirror placed at $45^{\degree}$ to the input light acting as the beamsplitter. In the $SU(1,1)$ interferometer, this beamsplitter is replaced by an optical parametric amplifier (OPA). This allows the squeezing of the light to occur in this active element within the interferometric setup. Experimental realization of the $SU(1,1)$ interferometer~\cite{SU11Expt1} has led to increased interest in exploiting the advantages of this active-squeezing-enhanced interferometry~\cite{SU11Expt2,SU11Expt3,SU11PRL,SU11StochPhase}. The results of~\cite{SU11StochPhase} indicate that the $SU(1,1)$ interferometer could surpass the precision of a conventional interferometer, when estimating the variance of a fluctuating phase. While there is considerable interest in introducing active elements in interferometric setups in various configurations to improve gravitational wave detection~\cite{IntSq1,IntSq2,IntSq3}, I confine the discussion in this paper to comparing the $SU(1,1)$ interferometer with the conventional $SU(2)$ interferometer. 

In this paper, I compare the precisions achieved by the $SU(1,1)$ interferometer and the conventional $SU(2)$ interferometer in estimating a fluctuating phase, especially in the context of spacetime fluctuations. I achieve this by computing the QFI of the variance of the fluctuating phase and the corresponding classical Fisher information (CFI) for different experimentally relevant measurement schemes. In both the $SU(2)$ and the $SU(1,1)$ interferometers, when the optical state at two input ports is a Gaussian state (i.e., a state with a Gaussian Wigner quasiprobability distribution), the state at the output ports remains Gaussian. Considering the experimental ease in generating Gaussian states as compared to any typical non-Gaussian state, I confine my investigation to Gaussian input states in my work. This allows me to use the prescription in \cite{QFI} to compute QFI using the covariance and mean of the $4$ quadrature observables corresponding to the bipartite state at the output ports. 

I also consider internal losses in both types of interferometers where a fraction of the light traversing the interferometer arms is lost. When applying experimentally relevant parameters, I find that the significantly small losses that are typical in a conventional $SU(2)$ interferometer, allow me to treat the interferometer essentially as a lossless interferometer. In contrast, considering current experimental setups, the $SU(1,1)$ interferometer suffers significant loss. Due to assumptions inherent to any theoretical modelling of an $SU(1,1)$ interferometer, constraints are placed on the maximum light intensity possible at the input ports. Accommodating such constraints and considering the extent of loss typical in \textit{current} experiments, I find that the $SU(1,1)$ interferometer does not offer any advantage over the $SU(2)$ interferometer in estimating the fluctuation strength and the correlation length of the spacetime fluctuations. However, possible future improvements in the $SU(1,1)$ interferometer could allow a wider range of parameter values. Considering such a possiblilty helps me identify parameter regimes at which the $SU(1,1)$ interferometer shows advantage over the $SU(2)$ interferometer.

The paper is structured as follows. In Sec. \ref{sec:models}, I compute the covariance matrix of the bipartite state at the output ports of the $SU(2)$ and $SU(1,1)$ interferometers in the presence of random spacetime fluctuations. In Sec. \ref{sec:res}, I find the QFI of the strength and correlation length of the spacetime fluctuations and the corresponding CFI both for homodyne and photon-number-counting measurements. I compare the Fisher information metrics between the two types of interferometers, to assess their relative advantages. In Sec. \ref{sec:conc}, I summarise and discuss the results.

\section{\label{sec:models} Models: $SU(2)$ and $SU(1,1)$ interferometers}

\begin{figure*}
\begin{tikzpicture}
\draw[red,thick] (0.,0.05) -- (1.95,0.05);
\draw[red,thick] (2.,0.) -- (4.5,0.);
\draw[red,thick] (2.,-2.) -- (2.,0.);
\draw[red,thick] (1.95,0.05) -- (1.95,2.);
\draw[red,thick,->] (0.,0.05) -- (0.6,0.05) node[above] {\color{black} $\aop{1}$};
\draw[red,thick,->] (2.,0.) -- (3.,0.);
\draw[red,thick,->] (1.95,0.05) -- (1.5,0.05) node[above] {\color{black} $\aop{3}$};
\draw[red,thick,->] (4.,0.) -- (3.5,0.);
\draw[red,thick,->] (2.,0.) -- (2.,-0.7) node[left] {\color{black} $\aop{4}$};
\draw[red,thick,->] (2.,-2.) -- (2.,-1.5) node[left] {\color{black} $\aop{2}$};
\draw[red,thick,->] (1.95,0.05) -- (1.95,1.); 
\draw[red,thick,->] (1.95,2.) -- (1.95,1.5);
\draw[black,thick] (4.5,-0.2) -- (4.5,0.2);
\draw[black,thick] (1.75,2.) -- (2.15,2.);
\draw[black,thick] (2.2,0.2) node[right] (scrip) {BS};
\draw[black!30,fill=black!30] (1.8,-0.2) -- (2.2,0.2)--(2.15,0.25)--(1.75,-0.15)--cycle;
\draw[black,thick] (1.8,-0.2) -- (2.2,0.2);
\draw[black,thick] (0.,0.05) node[left] (scrip) {A};
\draw[black,thick] (1.9,-2.) node[left] (scrip) {B};
\draw[black,thick] (4.5,0.) node[right] (scrip) {D};
\draw[black,thick] (2.,2.2) node[right] (scrip) {C};
\draw[black,thick,->] (0.2,-1.8) -- (0.5,-1.8) node[right] (scrip) {z};
\draw[black,thick,->] (0.2,-1.8) -- (0.2,-1.5) node[left] (scrip) {x};
\draw[black,thick,->] (0.2,-1.8) -- (0.1,-2) node[left] (scrip) {y};
\draw[black,thick] (-0.5,2.) node[left] (scrip) {(a)};
\end{tikzpicture}
\begin{tikzpicture}
\draw[red,thick] (0.,0.05) -- (1.95,0.05);
\draw[red,thick] (0.,0.15) -- (1.95,0.15);
\draw[red,thick] (2.,0.) -- (4.5,0.078);
\draw[red,thick] (4.5,0.078) -- (2.,0.156);
\draw[red,thick] (2.,-2.) -- (2.,0.);
\draw[red,thick] (2.15,-2.) -- (2.15,0.);
\draw[red,thick] (1.95,0.05) -- (2.028,2.);
\draw[red,thick] (2.028,2.) -- (2.106,0.05);
\draw[red,thick,->] (0.,0.05) -- (1.,0.05) node[below] {\color{black} $\aop{1}$};
\draw[red,thick,->] (1.95,0.15) -- (1.3,0.15) node[above] {\color{black} $\aop{3}$};
\draw[red,thick,->] (4.,0.098) -- (3.5,0.118);
\draw[red,thick,->] (2.15,0.) -- (2.15,-0.7) node[right] {\color{black} $\aop{4}$};
\draw[red,thick,->] (2.,-2.) -- (2.,-1.5) node[left] {\color{black} $\aop{2}$};
\draw[red,thick,->] (2.028,2.) -- (2.072,1.1);
\draw[black,thick] (4.5,-0.1) -- (4.5,0.3);
\draw[black,thick] (1.85,2.) -- (2.25,2.);
\draw[black!10,fill=black!10] (1.7,-0.3) -- (2.5,-0.3)--(2.5,0.5)--(1.7,0.5)--cycle;
\draw[black,thick] (1.7,-0.3)--(2.5,-0.3) -- (2.5,0.5)--(1.7,0.5)--cycle;
\draw[red,thick,snake it,->] (3.,1) node[right] {\color{black} Pump $\pi$-shifted} -- (2.5,0.5);
\draw[red,thick,->] (1.95,0.05) -- (1.988,1.);
\draw[red,thick,->] (2.,0.) -- (3.,0.04);
\draw[black!30,fill=black!30] (1.6,-0.4) -- (2.4,-0.4)--(2.4,0.4)--(1.6,0.4)--cycle;
\draw[black,thick] (1.6,-0.4)--(2.4,-0.4) -- (2.4,0.4)--(1.6,0.4)--cycle;
\draw[red,thick,snake it,->] (1,-1) node[left] {\color{black} Pump} -- (1.6,-0.4);
\draw[black,thick] (2.,0.) node (scrip) {OPA};
\draw[black,thick] (0.,0.05) node[left] (scrip) {A};
\draw[black,thick] (1.9,-2.) node[left] (scrip) {B};
\draw[black,thick] (4.5,0.1) node[right] (scrip) {D};
\draw[black,thick] (2.,2.2) node[right] (scrip) {C};
\draw[black,thick,->] (0.2,-1.8) -- (0.5,-1.8) node[right] (scrip) {z};
\draw[black,thick,->] (0.2,-1.8) -- (0.2,-1.5) node[left] (scrip) {x};
\draw[black,thick,->] (0.2,-1.8) -- (0.1,-2) node[left] (scrip) {y};
\draw[black,thick] (-0.5,2.) node[left] (scrip) {(b)};
\end{tikzpicture}
\begin{tikzpicture}
\draw[black,thick] (1,0.2) node (scrip) {BS/OPA};
\draw[black,thick] (1.7,-0.2) -- (0.3,-0.2)--(0.3,0.6)--(1.7,0.6)--cycle;
\draw[red,thick,->] (0.8,0.6) -- (0.8,1.); 
\draw[red,thick,->] (1.2,3.3) -- (1.2,2.8);
\draw[red,thick,dashed,->] (1.2,2.5) -- (2.3,2.5);
\draw[black,thick] (1.5,2.75) node[right] (scrip) {$\sqrt{\eta}$};
\draw[red,thick,->] (1.2,2.5) -- (1.2,2);
\draw[red,thick] (0.8,0.6) -- (0.8,3.3);
\draw[red,thick] (1.2,0.6) -- (1.2,3.3);
\draw[black,thick] (1.0,2.7) -- (1.4,2.3);
\draw[black,thick] (1.5,3.3) -- (0.5,3.3);
\draw[black,thick] (-0.5,3.5) node[left] (scrip) {(c)};
\end{tikzpicture}
\caption{A schematic diagram of an (a) $SU(2)$ and (b) $SU(1,1)$ interferometer. (c) Modelling internal loss in an interferometer with a beam splitter of reflectance $\sqrt{\eta}$.}
\label{fig:interferometer}
\end{figure*}
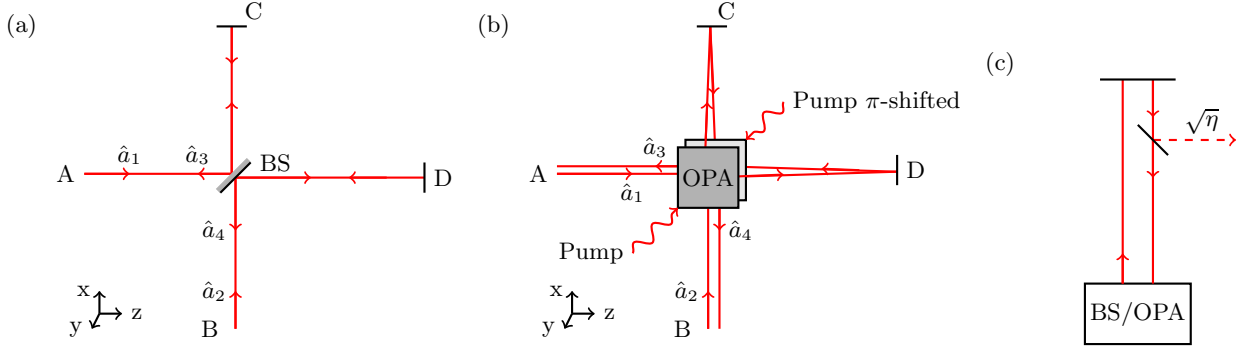

In this section, I find the covariance matrices of the bipartite state of the light at the output ports of the $SU(2)$ and the $SU(1,1)$ interferometers in the presence of random spacetime fluctuations. A schematic diagram of a conventional $SU(2)$ interferometer with a beamsplitter is shown in Fig. \ref{fig:interferometer} (a) and that of an $SU(1,1)$ interferometer with an OPA is shown in Fig. \ref{fig:interferometer} (b).

The light traversing the $\textsc{x}$-arm of any interferometer accumulates a phase $\Phi_{\textsc{x}}$ ($\textsc{x}=\textsc{c,d}$), which is given by~\cite{shar25},
\begin{align}
\Phi_{\textsc{x}} = \Phi_{\textsc{x} 0} + 2 \pi \frac{\Li}{\lambda} \int_{0}^{2} \rmd s' w(\bm{r}_{\textsc{x}}(s')).
\label{eqn:PhiDefn}
\end{align}
Here $\lambda=2 \pi c/\Omega$ is the wavelength of light with interferometer arm length $\Li$. The spacetime fluctuation (as defined in \cite{shar25}) at the spacetime point marked by the 4-vector $\bm{r}_{\textsc{x}}(s')$ along arm $\textsc{x}$ at the time instant $\Li s'/c$ is denoted by $w(\bm{r}_{\textsc{x}}(s'))$. Further, for a stationary, Gaussian noise, the ensemble average (denoted by the overline) over several realizations of spacetime fluctuations yields
\begin{align}
&\overline{w(\bm{r}_{\textsc{x}}(t))}=0,\\
&\overline{w(\bm{r}_{\textsc{x}}(t))\,w(\bm{r}_{\textsc{y}}(t'))}=\Gs \rho(\bm{r}_{\textsc{x}}(t)-\bm{r}_{\textsc{y}}(t')).
\end{align}
In what follows, I also use the approximation $\Phi_{\textsc{c} 0}=\Phi_{\textsc{d} 0}\equiv\Phi_{0}$ for analytical tractability. It is also useful to note that Eq. \eref{eqn:PhiDefn} is valid only in the long wavelength limit, as diffraction effects have been neglected. This implies that any length scale in the system is longer than the wavelength $\lambda$, including any characteristic correlation length of the spacetime fluctuations. 

Further, I define $s(q)=q $ if $q \leqslant 1$ and $s(q)=(2 - q)$ if $1<q\leqslant 2$ to describe the path light traverses in any interferometer arm. I also define the following correlation integrals.
\begin{widetext}
\begin{subequations}
\begin{eqnarray}
\nonumber \left(\frac{\lambda}{2 \pi \Li}\right)^{2}  \sigma &=&  \int\limits_{0}^{2} \rmd q_{1} \int\limits_{0}^{2} \rmd q_{2} \, \rho\left( \Li (q_{1}-q_{2}),0,0,\Li \left(s(q_{1})-s(q_{2})\right) \right)\\
&=& \int\limits_{0}^{2} \rmd q_{1} \int\limits_{0}^{2} \rmd q_{2} \, \rho\left(\Li (q_{1}-q_{2}),\Li \left(s(q_{1})-s(q_{2})\right),0,0\right),
\label{eqn:SigDefn}\\
\left(\frac{\lambda}{2 \pi \Li}\right)^{2}  \xi &=& \int\limits_{0}^{2} \rmd q_{1} \, \int\limits_{0}^{2} \rmd q_{2} \, \rho\left(\Li (q_{1}-q_{2}),\Li s(q_{1}),0,-\Li s(q_{2})\right).\label{eqn:XiDefn}
\end{eqnarray}
\label{eqn:CorrSet}
\end{subequations}
\end{widetext}
Here $\sigma$ and $\xi$ correspond to correlations within the same arm and across the two arms respectively.

The annihilation and creation operators of the optical modes at the two input ports of the any interferometer is denoted by $(\aop{1},\ad{1})$ and $(\aop{2},\ad{2})$ respectively, with $[\aop{i},\ad{i}]=1$ ($i=1,2$) and $[\aop{1},\ad{2}]=0$. The corresponding operators at the two output ports are $(\aop{3},\ad{3})$ and $(\aop{4},\ad{4})$ respectively. I denote the vector of quadrature operators of the optical modes at the output ports by $m={x_{3},p_{3},x_{4},p_{4}}$ with
\begin{align}
&x_{j}=\frac{\aop{j}+\ad{j}}{\sqrt{2}}, &p_{j}=\frac{\aop{j}-\ad{j}}{i \sqrt{2}}, (j=3,4).
\end{align}
The dispacement vector $D$ and the covariance matrix $C$ are given by
\begin{align}
D_{i}&=\overline{\bra{\psi} m_{i} \ket{\psi}}, \label{eqn:DispDefn} \\
C_{ij}&=\overline{\bra{\psi} m_{i} m_{j} \ket{\psi}} + \overline{\bra{\psi} m_{j} m_{i} \ket{\psi}} - 2 \overline{\bra{\psi} m_{i} \ket{\psi}} \hspace*{0.5 em} \overline{\bra{\psi} m_{j} \ket{\psi}}. \label{eqn:CovDefn}
\end{align}
Here  $\ket{\psi}$ is the bipartite state of the light at the input port. I find $D$ and $C$ for the $SU(2)$ and the $SU(1,1)$ interferometers in what follows.

\subsection{Lossless interferometers}
\subsubsection{$SU(2)$ interferometer}
The annihilation operators of the light at the output ports are written in terms of the corresponding operators at the input ports, as
\begin{align}
\begin{bmatrix}
\aop{3}\\
\aop{4}
\end{bmatrix} = \begin{bmatrix}
t & r\\
-r^{\ast} & t
\end{bmatrix} \begin{bmatrix}
e^{i \Phi_{\textsc{c}}} & 0\\
0 & e^{i \Phi_{\textsc{d}}}
\end{bmatrix} \begin{bmatrix}
t & r\\
-r^{\ast} & t
\end{bmatrix}^{\dagger} \begin{bmatrix}
\aop{1}\\
\aop{2}
\end{bmatrix},
\label{eqn:SU2trans}
\end{align}
where $|t|^{2}+|r|^{2}=1$ with $t=\cos \theta$, $r=\sin \theta e^{i \phi}$. I assume $\theta=\pi/4$ and $\phi=\pi/2$ for analytical tractability.

Simplifying Eq. \eref{eqn:SU2trans} yields
\begin{align}
\begin{bmatrix}
\aop{3}\\
\aop{4}
\end{bmatrix} = \begin{bmatrix}
\frac{e^{i \Phi_{\textsc{c}}} + e^{i \Phi_{\textsc{d}}}}{2} & - i \frac{(e^{i \Phi_{\textsc{c}}}-e^{i \Phi_{\textsc{d}}})}{2}\\
i \frac{(e^{i \Phi_{\textsc{c}}}-e^{i \Phi_{\textsc{d}}})}{2} & \frac{e^{i \Phi_{\textsc{c}}} + e^{i \Phi_{\textsc{d}}}}{2}
\end{bmatrix} \begin{bmatrix}
\aop{1}\\
\aop{2}
\end{bmatrix}.
\label{eqn:SU2simp}
\end{align}
I consider the state of light at the input ports to be $\ket{\psi}=\ket{\alpha}_{1} \otimes \ket{\beta}_{2}$ where $\ket{\alpha}_{1}$ denotes mode 1 being in the standard coherent state (assuming $\alpha \in \mathbb{R}$ for simplicity) and $\ket{\beta}_{2}$ denotes mode 2 being in a single-mode squeezed vacuum (assuming $\beta\in \mathbb{R}$  for simplicity). The squeezed state is defined as
\begin{align}
\ket{\beta}_{2}&=S(\beta) \ket{0}_{2},\\
\text{where, } \: S(\beta)&=\exp\left(\frac{\beta}{2}(\aopm{2}{2} - \adm{2}{2})\right),\\
\text{with } \: S^{\dagger}(\beta) \, \aop{2} \, S(\beta) &= \aop{2} \cosh \beta - \ad{2} \sinh \beta.
\label{eqn:sqeeze}
\end{align}
Further, I note that any non-zero $\Phi_{0}$ (recall $\Phi_{\textsc{c} 0}=\Phi_{\textsc{d} 0}\equiv\Phi_{0}$) in Eq. \eref{eqn:PhiDefn} can effectively be seen as a change in the arguments $\text{arg}(\alpha)$ and $\arg(\xi)$. Therefore, without loss of generality, I also set $\Phi_{0}=0$. 

Using Eqs. \eref{eqn:PhiDefn}-\eref{eqn:CovDefn} and \eref{eqn:SU2simp}-\eref{eqn:sqeeze}, I obtain the displacement vector and the covariance matrix in this case (see Appendix \ref{app:DispCov} for details). Empirically, it is seen that the fluctuation strength is small enough to assume $\alpha^{2}\gg (\sinh \beta)^{2}$, $\alpha^{2} \Gs \sigma \ll 1$, and neglect O($\Gs^{2}$). Retaining only the dominant terms, I find the dispacement vector
\begin{align}
D(\sigma)\approx\begin{pmatrix}
\sqrt{2} \, \alpha \, \left(1-\frac{\Gs \sigma}{2} \right)\\
0\\
0\\
0
\end{pmatrix},
\label{eqn:Disp2}
\end{align}
and the covariance matrix
\begin{align}
C^{(2)}\approx\begin{pmatrix}
1 & 0 & 0 & 0\\
0 & 1 + 2 (\js+\jc) & 0 & 0\\
0 & 0 & e^{-2 \beta} + 2 (\js-\jc) & 0 \\
0 & 0 & 0 & e^{2 \beta}
\end{pmatrix}.
\label{eqn:Cov2}
\end{align}
Here $\js=\alpha^{2} \Gs \sigma$ and $\jc=\alpha^{2} \Gs \xi$.

\subsubsection{$SU(1,1)$ interferometer}
For an $SU(1,1)$ interferometer,
\begin{align}
\begin{bmatrix}
\aop{3}\\
\ad{4}
\end{bmatrix} = \begin{bmatrix}
\mu & \nu\\
\nu^{\ast} & \mu^{\ast}
\end{bmatrix} \begin{bmatrix}
e^{i \Phi_{\textsc{c}}} & 0\\
0 & e^{-i \Phi_{\textsc{d}}}
\end{bmatrix} \begin{bmatrix}
\mu & \nu\\
\nu^{\ast} & \mu^{\ast}
\end{bmatrix}^{-1} \begin{bmatrix}
\aop{1}\\
\ad{2}
\end{bmatrix}.
\label{eqn:SU11trans}
\end{align}
Here $|\mu|^{2}-|\nu|^{2}=1$ and I assume $\mu=\cosh \beta$, $\nu=\sinh \beta$ ($\beta \in \mathbb{R}$) for analytical tractability.

Simplifying Eq. \eref{eqn:SU11trans} yields
\begin{widetext}
\begin{align}
& \begin{bmatrix}
\aop{3}\\
\ad{4}
\end{bmatrix} = \begin{bmatrix}
(\cosh \beta)^{2} \, e^{i \Phi_{\textsc{c}}} - (\sinh \beta)^{2} \, e^{-i \Phi_{\textsc{d}}} & \frac{ (e^{-i \Phi_{\textsc{d}}}-e^{i \Phi_{\textsc{c}}}) \, \sinh 2 \beta}{2}\\
\frac{(e^{i \Phi_{\textsc{c}}}-e^{-i \Phi_{\textsc{d}}})\,\sinh 2 \beta }{2} & -(\sinh \beta)^{2} \,e^{i \Phi_{\textsc{c}}} + (\cosh \beta)^{2} \, e^{-i \Phi_{\textsc{d}}}
\end{bmatrix}\begin{bmatrix}
\aop{1}\\
\ad{2}
\end{bmatrix}.
\label{eqn:SU11simp}
\end{align}
\end{widetext}
The state at the input ports is $\ket{\psi}=\ket{\alpha}_{1} \otimes \ket{0}_{2}$ where $\ket{0}_{2}$ denotes mode 2 being in the vacuum state. As the state is subject to squeezing at the OPAs within the interferometer, the input state is left unsqueezed.

As in the case of the $SU(2)$ interferometer, I use Eqs. \eref{eqn:PhiDefn}-\eref{eqn:CovDefn} and \eref{eqn:SU11simp} to obtain the displacement vector and covariance matrix, retaining only the dominant terms. I find that the dispacement vector is the same as that given in Eq. \eref{eqn:Disp2}. The covariance matrix
\begin{align}
C^{(11)}\approx\begin{pmatrix}
1 & 0 & 0 & 0\\
0 & 1 +\chi_{1} + \chi_{2} & 0 & \chi_{3}\\
0 & 0 & 1 & 0 \\
0 & \chi_{3} & 0 & 1+\chi_{2}
\end{pmatrix},
\label{eqn:Cov11}
\end{align}
where
\begin{align}
\chi_{1}&=4 \, \js,\\
\chi_{2}&=2 \, (\sinh 2 \beta)^{2} \, (\js+\jc),\\
\chi_{3}&=-\sinh 4 \beta \, (\js+\jc).
\end{align}

\subsection{Lossy interferometers}
I consider equal loss in the two arms of any interferometer with the ratio of light lost out of each arm is denoted by $\eta$ (modelled as in Fig. \ref{fig:interferometer} (c)).

\subsubsection{$SU(2)$ interferometer}
The dispalcement vector $D_{\textsc{l}}$ and the covariance matrix $C^{(2)}_{\textsc{l}}$ in the presence of internal losses are
\begin{align}
&D_{\textsc{l}}=\sqrt{(1 - \eta)} \, D, \label{eqn:Disp2Lossy}\\
&C^{(2)}_{\textsc{l}}=\eta \, \mathbb{I}_{4} + (1-\eta) \, C^{(2)},
\label{eqn:Cov2Lossy}
\end{align}
where $\mathbb{I}_{4}$ is a $4\times4$ identity matrix.

\subsubsection{$SU(1,1)$ interferometer}
The covariance matrix with loss
\begin{widetext}
\begin{align}
C^{(11)}_{\textsc{l}}= \eta \, \begin{pmatrix}
\cosh 2 \beta & 0 & \sinh 2 \beta & 0\\
0 & \cosh 2 \beta & 0 & -\sinh 2 \beta\\
\sinh 2 \beta & 0 & \cosh 2 \beta & 0 \\
0 & -\sinh 2 \beta & 0 & \cosh 2 \beta
\end{pmatrix} + (1-\eta) \, C^{(11)}.
\label{eqn:Cov11Lossy}
\end{align} 
\end{widetext}

\section{\label{sec:res} Results}
In this section, I find the classical and quantum Fisher information metrics for estimating two characteristic parameters of the spacetime fluctuations, namely, the fluctuation strength $\Gs$ and the correlation length $\ell$ specific to a given two-point correlation function $\rho$. I also compare the QFI with the corresponding CFI for different experimentally relevant measurement schemes for the two types of interferometers. This allows me to assess the advantages and limitations of the two types of interferometers in estimating the parameters of interest.

As the states of interest are Gaussian, I use the procedure prescribed in \cite{QFI} to obtain the QFI in terms of any given covariance matrix $C$ and displacement vector $D$. The QFI for estimating any parameter $\zeta$ is 
\begin{align}
\nonumber Q_{\zeta}=& \frac{1}{2} \left(\rmd_{\zeta} \vec{C}\right)^{T} (C \otimes C - M \otimes M)^{-1} \left(\rmd_{\zeta} \vec{C}\right)\\
& + 2 \left(\rmd_{\zeta} D\right)^{T} C^{-1} \left(\rmd_{\zeta} D\right) ,
\label{eqn:QFI}
\end{align}
where $\rmd_{\zeta} O \equiv \frac{\rmd 0}{\rmd \zeta}$ for any $O$ that is a function of $\zeta$, $\vec{C}$ is the vectorised covariance matrix, and the commutator of the quadrature operators $i M_{ij}= [m_{i},m_{j}]$. Recall that the vector of quadrature operators of the optical modes at the output ports $m={x_{3},p_{3},x_{4},p_{4}}$.

In both the interferometers considered, a measurement of the observable $m_{i}$ yields a Gaussian probability distribution with mean $D_{i}$ and variance $C_{ii}$ ($i=1,2,3,4$). The CFI corresponding to such a Gaussian probability distribution is
\begin{align}
C^{(\textsc{g}i)}_{\zeta}=\frac{1}{2} \left(\frac{\rmd_{\zeta} C_{ii}}{C_{ii}}\right)^{2} + 2  \,\left[\frac{(\rmd_{\zeta} D_{i})^{2}}{C_{ii}}\right].
\label{eqn:CFIhomo}
\end{align} 

A photon number measurement typically yields a non-Gaussian probability distribution. Denoting the probability of measuring $j$-photons by $p_{j}$ in any given port, the CFI is given by~\cite{QFIparis}
\begin{align}
C^{(\textsc{n}n)}_{\zeta}= \sum_{j=0}^{n} \frac{(\rmd_{\zeta} p_{j})^{2}}{p_{j}} + \frac{(\rmd_{\zeta} Q)^{2}}{Q}
\label{eqn:CFIPhNum}
\end{align}
where $n$ is the maximum number of photons upto which the photons are being counted discretely and $Q=1-\sum_{j=0}^{n} p_{j}$. It is also important here to list the steps necessary for obtaining $p_{j}$ in port $i$. I find the Wigner function $W(Z)$ for any $2$-dimensional vector $Z$, in terms of the covariance matrix $C^{(i)}$ and the displacement vector $D^{(i)}$ corresponding to the state at the output port $i$, as
\begin{align}
W\left(Z\right) = \frac{\exp \left( - (Z-D^{(i)})^{T} (C^{(i)})^{-1} (Z-D^{(i)}) \right)}{\pi \sqrt{\text{det} \, C^{(i)}}}.
\end{align}
Here $\text{det}\, C^{(i)}$ denotes the determinant of $C^{(i)}$. The probability
\begin{align}
\nonumber p_{j}=2 (-1)^{j} \int_{-\infty}^{\infty} &\rmd Z_{1} \int_{-\infty}^{\infty} \rmd Z_{2} \exp \left(-Z_{1}^{2}-Z_{2}^{2}\right)\\
& L_{j}\left(2\left(Z_{1}^{2}+Z_{2}^{2}\right)\right) W\left(Z_{1},Z_{2}\right),
\end{align}
where $L_{j}(x)$ is the Laguerre polynomial.

Using Eqs. \eref{eqn:Disp2Lossy}-\eref{eqn:Cov11Lossy} in Eqs. \eref{eqn:QFI}-\eref{eqn:CFIPhNum}, I compute the Fisher information metrics for estimating the fluctuation strength $\Gs$, namely, QFI $Q_{\Gs}$ and the CFIs $C^{(\textsc{g}i)}_{\Gs}$ ($i=1,2,3,4$) and $C^{(\textsc{n}n)}_{\Gs}$ in the case of $SU(2)$ and $SU(1,1)$ interferometers. I scale the Fisher information metrics by $\Gs^{2}$, for ease of comparison. I recall that I have assumed $\alpha^{2}\gg (\sinh \beta)^{2}$, $\js=\alpha^{2} \Gs \sigma \ll 1$, and considered to leading order in $\Gs$. Here, I also assume $\sigma \gg \xi$ as I expect correlations within an arm to be greater than that across two arms of an interferometer. I find that the scaled QFI for either type of interferometers is
\begin{align}
\Qs= \Gs^{2} Q_{\Gs} = 
\begin{cases} 
(1-\eta) \js, \: &\text{for } \beta\to 0\\
\frac{1}{2}, \: &\text{for } e^{-2 \beta} \ll (1-\eta) \js
\end{cases}
\label{eqn:QGs}
\end{align}

In the case of an $SU(2)$ interferometer, the CFI for measuring quadrature observables is maximum for $m_{2}$ and $m_{3}$. The scaled CFI corresponding to $m_{3}$ is
\begin{align}
\Cg^{(3)}=\Gs^{2} C^{(\textsc{g}3)}_{\Gs}=
\begin{cases}
2 \, (1-\eta)^{2} \, \js^{2}, \: &\text{for } \beta\to 0\\
\frac{1}{2}, \: &\text{for } e^{-2 \beta} \ll (1-\eta) \js
\end{cases}
\label{eqn:Cg2Gs2}
\end{align}
with $\Cg^{(2)}=2 \, (1-\eta)^{2} \, \js^{2}$ for any $\beta$. As $\left(\rmd_{\Gs}C^{(2)}_{\textsc{l}}\right)^{-1}$ is diagonal and positive in an $SU(2)$ interferometer, I know that the photon counting at either port is the optimal measurement~\cite{QFI}. This is illustrated by the CFI for a photon number measurement with $n=1$ at output mode labelled as 4, or equivalently, at the dark port. The CFI is
\begin{align}
\lim_{\beta\to 0} \Cn^{(1)}= \lim_{\beta\to 0} \Gs^{2} C^{(\textsc{n}1)}_{\Gs} = \frac{1}{2} (1-\eta) \js = \frac{1}{2} \lim_{\beta\to 0} \Qs.
\label{eqn:CnGs2}
\end{align}
This is evidently the optimal measurement in the limit $\beta\to0$ because the QFI contribution from the dark port is $\frac{1}{2} \Qs$ in this limit. As $\beta$ increases, the optimal measurement remains to be the optimal measurement. However, as the number of photons at the dark port increases by $(\sinh \beta)^{2}$, a dicrete measurement of photons to increasing $n$ is needed. The CFI for a photon number measurement at the bright port is also optimal but the $n$ to which the photons need to be measured discretely is decided by both $\alpha^{2}$ and $\beta$.

In the case of the $SU(1,1)$ interferometer, the CFI for measuring quadrature observables is maximum for $m_{2}$ and $m_{4}$. The scaled CFI corresponding to $m_{2}$ is
\begin{align}
\Cg^{(2)}=\Gs^{2} C^{(\textsc{g}2)}_{\Gs}=
\begin{cases}
8 (1-\eta)^{2} \js^{2}, \: &\text{for } \beta\to 0\\
\frac{1}{2}, \: &\text{for } e^{-2 \beta} \ll (1-\eta)\js.
\end{cases}
\label{eqn:Cg2Gs11}
\end{align}
The scaled CFI corresponding to $m_{4}$ is
\begin{align}
\Cg^{(4)}=\Gs^{2} C^{(\textsc{g}4)}_{\Gs}=
\begin{cases}
0, \: &\text{for } \beta\to 0\\
\frac{1}{2}, \: &\text{for } e^{-2 \beta} \ll (1-\eta)\js.
\end{cases}
\label{eqn:Cg4Gs11}
\end{align}
As shown in \cite{QFI}, photon counting in suitably defined modes is the optimal measurement if $\left(\rmd_{\zeta}C\right)^{-1}$ has a definite signature. Though $\left(\rmd_{\Gs}C^{(11)}_{\textsc{l}}\right)^{-1}$ is positive, the optimal measurement requires discrete measurement of photon numbers up to an $n$ decided by both $\alpha^{2}$ and $\beta$, which would typically make it inconveniently large.

\begin{figure*}
\includegraphics[width=\textwidth]{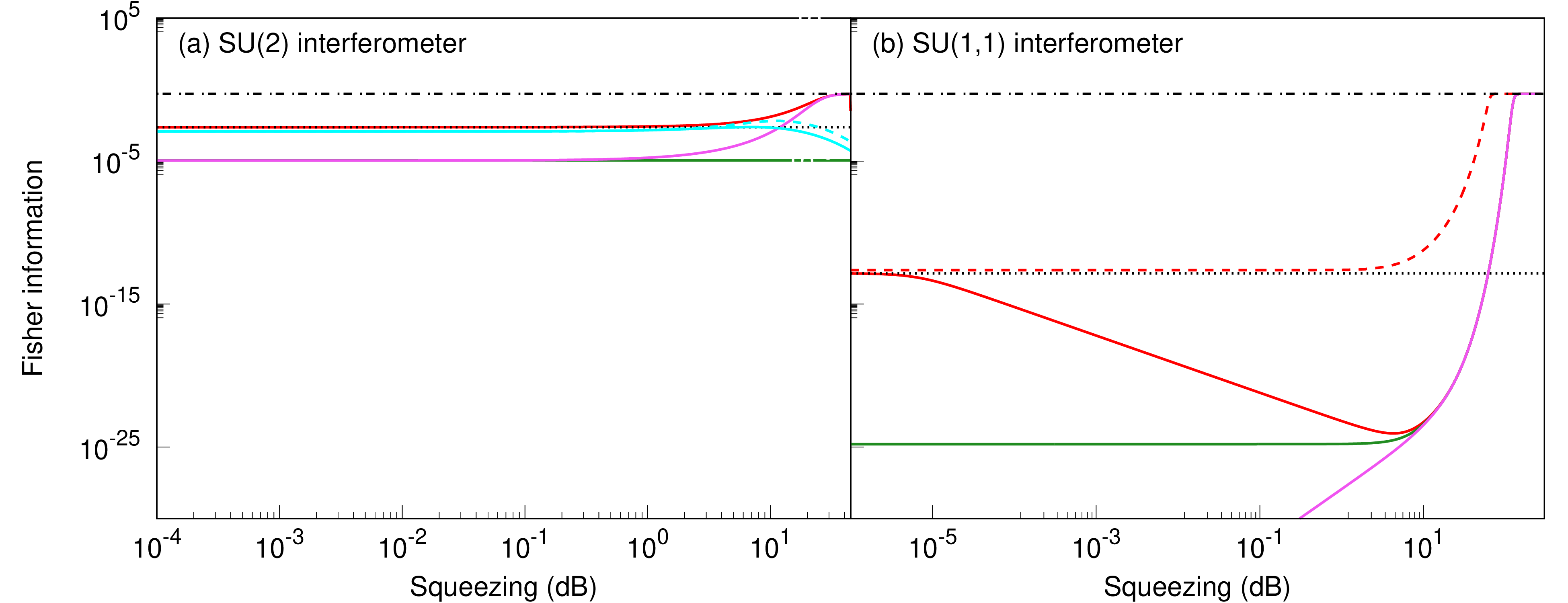}
\caption{Scaled QFI $\Qs=\Gs^{2} Q_{\Gs}$ (red), scaled CFI $\Cg^{(2)}=\Gs^{2} C^{(\textsc{g}2)}_{\Gs}$ (green), scaled CFI $\Cg^{(i)}=\Gs^{2} C^{(\textsc{g}i)}_{\Gs}$ (magenta) with $i=3$ in (a) and $i=4$ in (b), and scaled CFI $\Cn^{(n)}=\Gs^{2} C^{(\textsc{n}n)}_{\Gs}$ with $\Cn^{(1)}$ (solid cyan), $\Cn^{(5)}$ (dashed cyan) vs squeezing $10 \log_{10}\left(e^{2 \beta}\right)$ in decibels for (a) $SU(2)$ and (b) $SU(1,1)$ interferometers. The black horizontal lines are at 0.5 (dot-dashed) and $(1-\eta)\,\js$ (dotted) with $\eta=5\times10^{-6}$ (respectively, $\eta=0.4$) in an $SU(2)$ (resp., $SU(1,1)$) interferometer. The red dashed curve in (b) plots scaled QFI $\Qs=\Gs^{2} Q_{\Gs}$ in the lossless case, for comparison. (A magnified figure of the left panel is present in Appendix \ref{app:SU2FI} for the sake of clarity).}
\label{fig:QGs}
\end{figure*}

I illustrate these results in Fig. \ref{fig:QGs}. For numerical computation, I consider
\begin{subequations}
\begin{align}
\sigma &=\left(\frac{2 \pi}{\lambda} \right)^{2} 2 \ell (\Li-\ell) ,\\
\xi &= \left(\frac{2 \pi}{\lambda} \right)^{2} \pi \ell^{2},
\end{align}
\label{eqn:sigxi}
\end{subequations}
corresponding to a correlation function $\rho(\bm{r}_{1}-\bm{r}_{2})= e^{-\frac{\Vert \vec{r}_{1}-\vec{r}_{2} \Vert}{\ell}} \Theta(\Vert \vec{r}_{1}-\vec{r}_{2} \Vert - c |t_{1}-t_{2}|)$ in the limit $\Li\gg \ell$. Here $\ell$ is the correlation length, $\Li$ is the interferometer arm length, and $\lambda$ is the wavelength of the light traversing in the interferometers. I also assume the following values for the different parameters, first considering current experimentally achievable systems. I use $\alpha=10^{10}$, $\eta=5\times10^{-6}$, and $\lambda=10^{-6}$ m for an $SU(2)$ interferometer as in current setups such as QUEST~\cite{QUESTdata} and GQuEST~\cite{smv24}. In contrast, I consider $\alpha=10^{4}$, $\eta=0.4$, and $\lambda=10^{-7}$ m in an $SU(1,1)$ interferometer to conform with experimenets reported in \cite{SU11Expt3,SU11Expt4}. The chosen $\alpha$ value allows the lossy $SU(1,1)$ interferometer to have an undepeleted pump and identical gain in both the OPAs, with perfect mode matching. In both types of interferometers, I consider the interferometer arm length $\Li=3$ m and I assume the spacetime fluctuations to have a strength $\Gs=10^{-35}$ and correlation length $\ell=0.01$ m.

In Fig. \ref{fig:QGs} (a), the Fisher information metrics scaled by $\Gs^{2}$ is plotted as a function of the extent of squeezing in the state at the input port of the $SU(2)$ interferometer. The results listed above are clearly illustrated. Further, as the loss $\eta\ll1$ in any experimentally relevant $SU(2)$ interferometer, both QFI and CFI are almost identical to the lossless case. I also note that the results obtained estimating $\Gs$ in an $SU(2)$ interferometer can be compared to the results obtained by estimating the variance (instead of standard deviation) of the spacetime fluctuations in \cite{gardner2025}. A straightforward comparison can be made in the lossless case using Eq. (33) in \cite{gardner2025}. By equating the variance in \cite{gardner2025} to $\js$ (a valid approximation when considering $\sigma \gg \xi$), I can see that the QFIs in the lossless case agree with those obtained in my work. However, in my work, I explicitly show how the QFI is enhanced by $\alpha^{2}$ and also obtain the condition $e^{-2 \beta} \ll (1-\eta) \, \js$ that is needed to attain maximum QFI.
 
In Fig. \ref{fig:QGs} (b), as before, the Fisher information metrics scaled by $\Gs^{2}$ is plotted as a function of the extent of squeezing generated by the OPAs in the $SU(1,1)$ interferometer. The effect of loss in each arm is significant, as the loss $\eta$ is also significantly larger in the $SU(1,1)$ interferometer compared to the conventional interferometers. To illustrate this, the $\Qs$ in the lossless case is plotted along with $\Qs$ corresponding to the lossy interferometer. When considering experimentally relevant parameter values, it is immediately evident that the $SU(1,1)$ interferometer does not provide any advantage over the conventional interferometers in estimating the fluctuation strength of the spacetime fluctuations.

\begin{figure*}
\includegraphics[width=\textwidth]{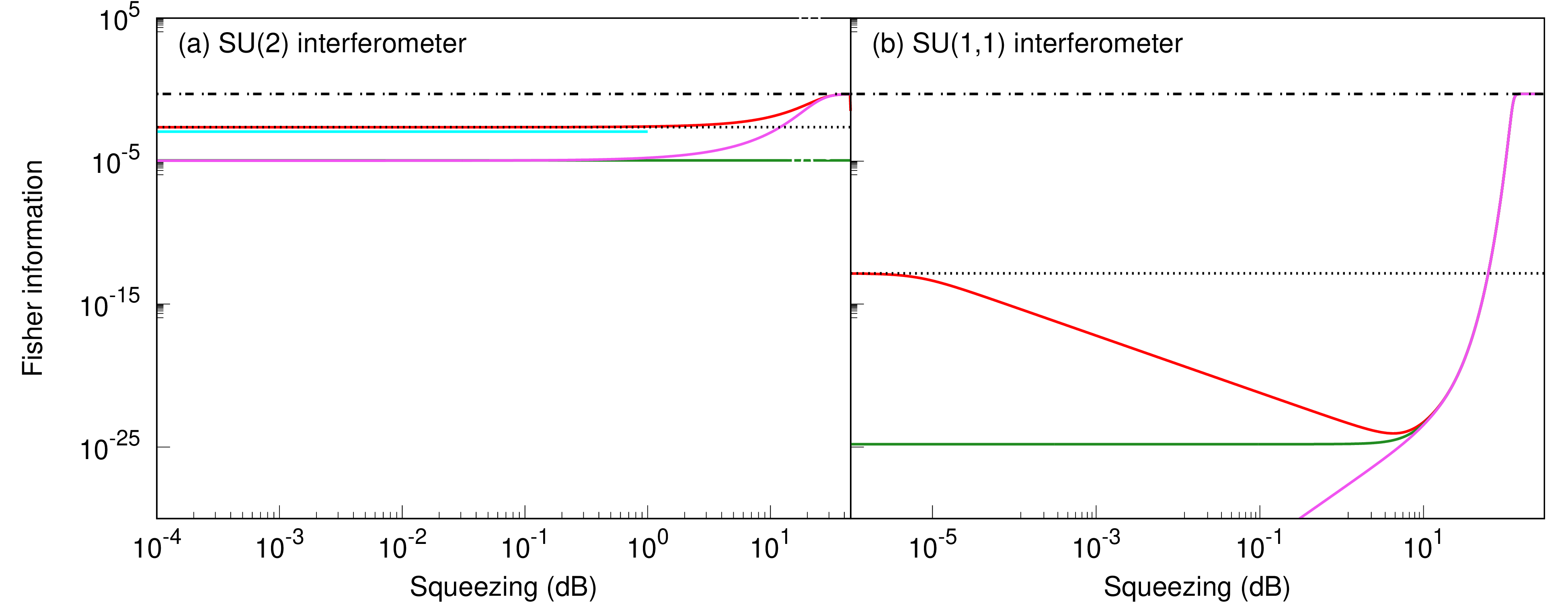}
\caption{Scaled QFI $\Qc=\ell^{2} Q_{\ell}$ (red), scaled CFI $\Cgc^{(2)}=\ell^{2} C^{(\textsc{g}2)}_{\ell}$ (green), scaled CFI $\Cgc^{(i)}=\ell^{2} C^{(\textsc{g}i)}_{\ell}$ (magenta) with $i=3$ in (a) and $i=4$ in (b), and scaled CFI $\Cnc^{(n)}=\ell^{2} C^{(\textsc{n}n)}_{\ell}$ with $\Cnc^{(1)}$ (cyan) vs squeezing $10 \log_{10}\left(e^{2 \beta}\right)$ in decibels for (a) $SU(2)$ and (b) $SU(1,1)$ interferometers. The black horizontal lines are at 0.5 (dot-dashed) and $(1-\eta)\,\js$ (dotted) with $\eta=5\times10^{-6}$ (respectively, $\eta=0.4$) in an $SU(2)$ (resp., $SU(1,1)$) interferometer.}
\label{fig:Qlr}
\end{figure*}

If $\sigma$ and $\xi$ are of the form given in Eq.~\eref{eqn:sigxi}, I find that the QFI for estimating $\ell$ in both types of interferometers is
\begin{align}
\Qc= \ell^{2} Q_{\ell} = 
\begin{cases} 
(1-\eta) \js, \: &\text{for } \beta\to 0\\
\frac{1}{2}, \: &\text{for } e^{-2 \beta} \ll (1-\eta) \js.
\end{cases}
\end{align}
I note that this is identical to the scaled QFI $\Qs$ for estimating $\Gs$ in Eq. \eref{eqn:QGs}. I similarly find that the scaled CFIs $\Cgc^{(i)}=\ell^{2} C^{(\textsc{g}i)}_{\ell}$ and $\Cnc^{(n)}=\ell^{2} C^{(\textsc{n}n)}_{\ell}$ for estimating $\ell$ are also identical to $\Cg^{(i)}$ (Eqs. \eref{eqn:Cg2Gs2}, \eref{eqn:Cg2Gs11} and \eref{eqn:Cg4Gs11}) and $\Cn^{(n)}$ (Eq. \eref{eqn:CnGs2}) respectively. This is illustrated in Fig. \ref{fig:Qlr}. 

\begin{figure}
\includegraphics[width=0.5\textwidth]{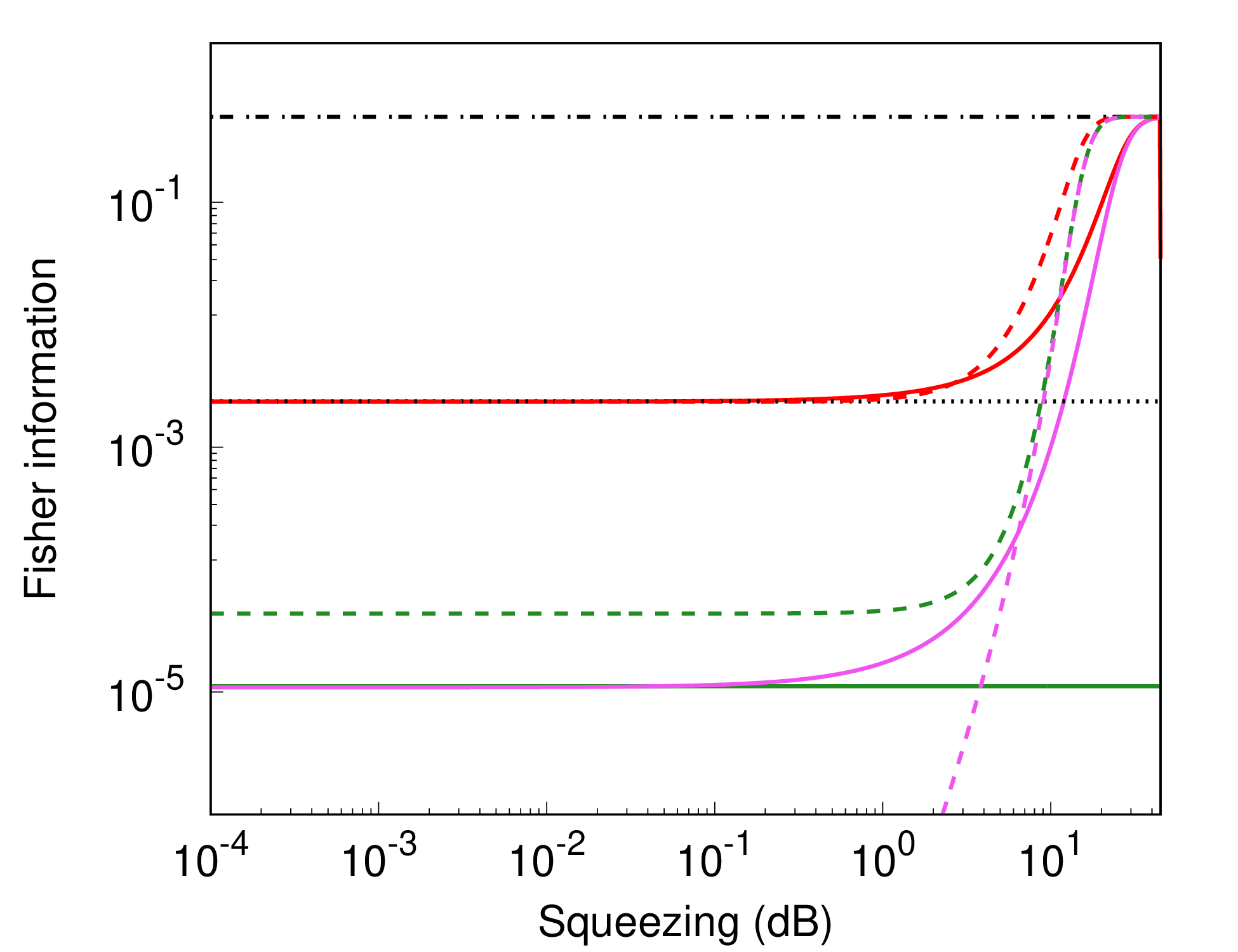}
\caption{Scaled QFI $\Qs=\Gs^{2} Q_{\Gs}$ (red), scaled CFI $\Cg^{(2)}=\Gs^{2} C^{(\textsc{g}2)}_{\Gs}$ (green), and $\Cg^{(i)}=\Gs^{2} C^{(\textsc{g}i)}_{\Gs}$ (magenta) vs squeezing $10 \log_{10}\left(e^{2 \beta}\right)$ in decibels, with solid curves for $SU(2)$ and dashed curves for $SU(1,1)$ interferometers. Here $i=3$ for $SU(2)$ and $i=4$ for $SU(1,1)$ interferometers. The black horizontal lines are at 0.5 (dot-dashed) and $(1-\eta)\,\js$ (dotted) with $\eta=5\times10^{-6}$.}
\label{fig:FIcomp}
\end{figure}

However, it is reasonable to consider possible future developments in $SU(1,1)$ interferometry, which could allow higher light intensity than that allowed in current setups. Considering such possible developments, I assume identical parameter values in both types of interferometers: $\alpha=10^{10}$, $\eta=5\times10^{-6}$, and $\lambda=10^{-6}$ m. As before, $\Li=3$ m, $\Gs=10^{-35}$ and $\ell=0.01$ m. It is evident from Fig. \ref{fig:FIcomp}, that in such a case, $SU(1,1)$ interferometer shows advantage over the $SU(2)$ interferometer. Figure \ref{fig:FIcomp} illustrates this result in the case of estimating $\Gs$. I note that this result holds true for estimating $\ell$ too. This is also illustrated using Table \ref{tab:comp} in which the ratio of any given Fisher information metric of the $SU(1,1)$ interferometer to that of the $SU(2)$ interferometer is reported. Here the ratio of the CFIs is taken considering the most efficient measurement of a quadrature obserable in each case. The ratios of the CFIs being greater than 1 illustrates that a real advantage of the $SU(1,1)$ interferometer can be realised with suitably chosen homodyne measurements.
  
\begin{table}
\begin{tabular}{ccccc}
\hline
\hline
Squeezing & $\Qs$ ratio & $\Cg$ ratio  & $\Qc$ ratio & $\Cgc$ ratio\\
\hline
\hline
6 dB & 1.790 & 1.899 & 1.817 & 1.932\\
10 dB & 4.276 & 6.200 & 4.357 & 6.327 \\
\hline
\end{tabular}
\caption{Comparing the ratio of Fisher information metrics corresponding to $SU(1,1)$ interferometer to that corresponding to $SU(2)$ interferometer.}
\label{tab:comp}
\end{table}

\section{\label{sec:conc} Discussion}

I consider two types of interferometric setups: the conventional $SU(2)$ interferometer, and the $SU(1,1)$ interferometer with two OPAs. To be experimentally relevant, I consider Gaussian states at the input ports of both interferometers. Assuming a stationary, Gaussian, random process to model spacetime fluctuations, I compute the displacement vector and the covariance matrices of the quadrature variables of the optical modes at the output ports, both in the presence and absence of internal loss in the interferometers. Using this, I compute the Fisher information metrics in both types of interferometers. Considering the weakness of the fluctuation strength of the spacetime fluctuations, the Fisher information metrics are computed only to leading order in the fluctuation strength. I compare the Fisher information metrics corresponding to the $SU(1,1)$ interferometer with that of the $SU(2)$ interferometer. As should be expected, I find that, on considering parameter values relevant to \textit{current} experiments, the $SU(1,1)$ interferometer does not provide any advantage over the conventional interferometers in estimating either the fluctuation strength or the correlation length of the spacetime fluctuations. However, it is possible to harness a real advantage with improved technology in $SU(1,1)$ interferometry. Comparing Fisher information metrics corresponding to this improved $SU(1,1)$ interferometer with that of the conventional $SU(2)$ interferometer shows a clear advantage, that can be realised by a suitable homodyne measurement at the output port.

However, it is important to highlight some caveats. Considering the weakness of the spacetime fluctuations, $\js$ is expected to be much smaller than $e^{-2\beta}$ typically, making the low squeezing limit more relevant. The QFI is identical in form in both types of interferometers in the low squeezing limit, leading to the $SU(1,1)$ interferometer carrying no significant advantage in this limit. Further, in the $SU(2)$ interferometer, the photon number measurement at the dark port offers an optimal measurement where one needs to measure discretely only up to a single photon. In contrast, I find no such experimentally convenient and optimal measurement scheme in the case of the $SU(1,1)$ interferometer at the same limit. This indicates the challenges in harnessing advantage from the $SU(1,1)$ interferometer, even on improving the characteristics of OPA to allow higher number of photons in the signal. To harness any real advantage of the $SU(1,1)$ interferometer, a large gain factor needs to be achieved in addition to improving the OPA to allow higher signal intensity.

In addition to identifying the parameter regimes at which a real advantage of the $SU(1,1)$ interferometer is achieved, this work also extends the analysis of \cite{gardner2025} in the context of the conventional interferometer. By modelling both the output modes of the interferometer, I find that the QFI is enhanced by the number of photons in the input coherent state. I also obtain the extent of squeezing necessary, $e^{-2 \beta} \ll (1-\eta) \, \js$, to attain maximum QFI. This work establishes what is necessary to maximise the information gleaned from existing interferometric setups such as QUEST~\cite{QUESTdata} and GQuEST~\cite{smv24}, while also showing what will be needed in possible futuristic $SU(1,1)$ interferometers to surpass the sensitivity of current $SU(2)$ interferometers.
 
\begin{acknowledgments}
I acknowledge the support of the Leverhulme Trust under research grant ECF-2024-124. I also thank Prof Animesh Datta for extensive discussions that were crucial for this work.

\end{acknowledgments}


%

\appendix
\section{\label{app:DispCov} Computing displacement vector and covariance matrices}
I illustrate the steps involved in obtaining the covariance matrix $C$ and the displacement vector $D$ by listing the steps involved in obtaining the expection value of any observable that is a function of the form $f(\aop{3},\ad{3},\aop{4},\ad{4})$ in either type of interfeometer. This function $f$ is mapped on to $\sum_{k} c_{k}(\Phi_{\textsc{c}},\Phi_{\textsc{d}}) \widetilde{f}_{k}(\aop{1},\ad{1},\aop{2},\ad{2})$, using Eq. \eref{eqn:SU2simp} (resp., Eq. \eref{eqn:SU11simp}) for an $SU(2)$ (resp., $SU(1,1)$) interferometer. 

I obtain
\begin{align}
\nonumber &\overline{\bra{\psi} f(\aop{3},\ad{3},\aop{4},\ad{4}) \ket{\psi}}\\
&=\sum_{k} \overline{c_{k}(\Phi_{\textsc{c}},\Phi_{\textsc{d}})} \bra{\psi}\widetilde{f}_{k}(\aop{1},\ad{1},\aop{2},\ad{2})\ket{\psi},
\end{align}
where $\ket{\psi}$ is the bipartite state of the light at the input port. Considering that the noise is Gaussian, I find closed form expressions when computing $\overline{c_{k}(\Phi_{\textsc{c}},\Phi_{\textsc{d}})}$ and I retain only upto first-order in fluctuation strength $\Gamma$.

\section{\label{app:SU2FI} Fisher information metrics for estimating $\Gs$ in the $SU(2)$ interferometer}

I present the magnified version of the left panel of the Fig. \ref{fig:QGs}, for the sake of clarity.
 
\begin{figure}
\includegraphics[width=0.5\textwidth]{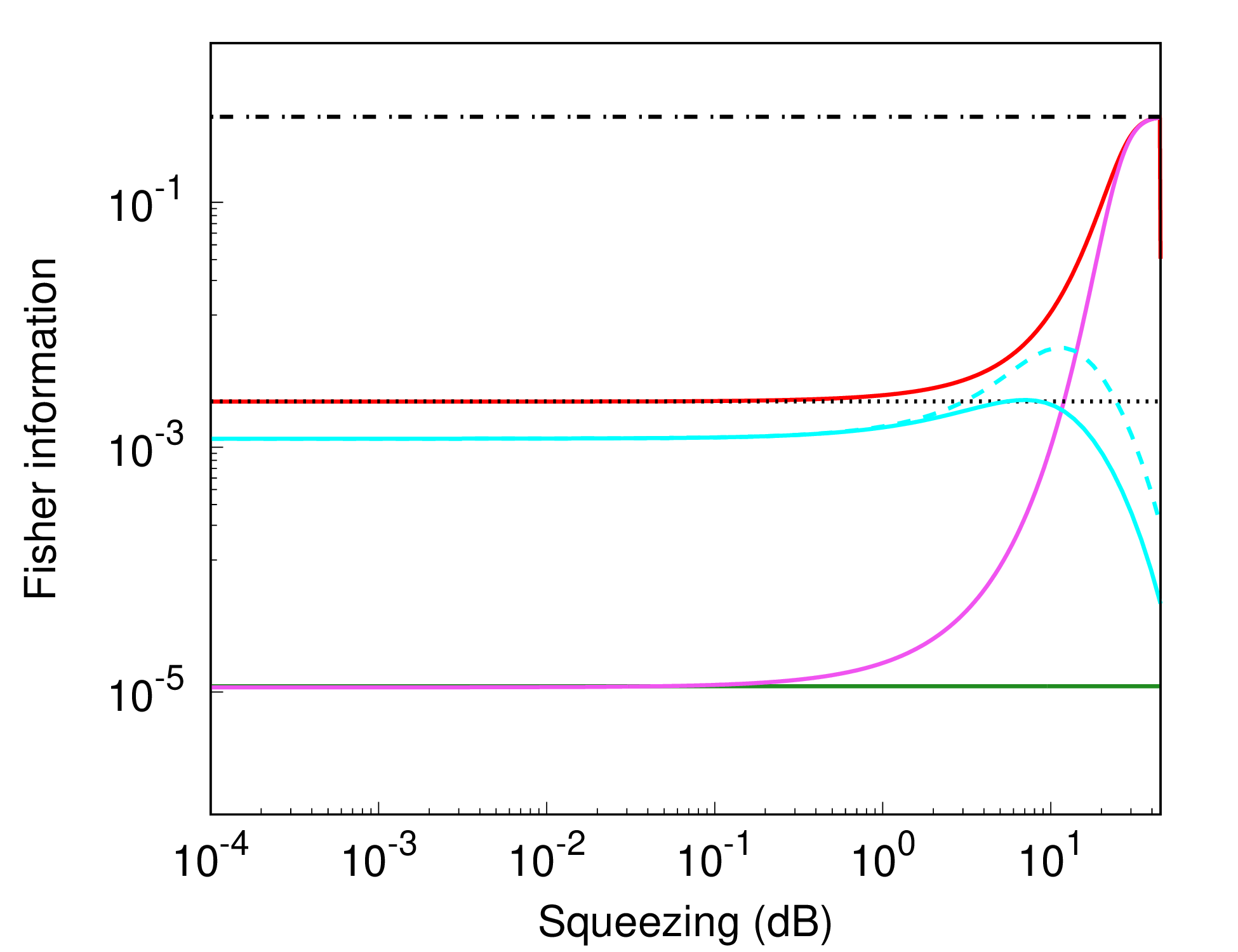}
\caption{Scaled QFI $\Qs=\Gs^{2} Q_{\Gs}$ (red), scaled CFI $\Cg^{(2)}=\Gs^{2} C^{(\textsc{g}2)}_{\Gs}$ (green), scaled CFI $\Cg^{(3)}=\Gs^{2} C^{(\textsc{g}3)}_{\Gs}$ (magenta) and scaled CFI $\Cn^{(n)}=\Gs^{2} C^{(\textsc{n}n)}_{\Gs}$ with $\Cn^{(1)}$ (solid cyan), $\Cn^{(5)}$ (dashed cyan) vs squeezing $10 \log_{10}\left(e^{2 \beta}\right)$ in decibels for an $SU(2)$ interferometer. The black horizontal lines are at 0.5 (dot-dashed) and $(1-\eta)\,\js$ (dotted) with $\eta=5\times10^{-6}$.}
\end{figure}

\bibliography{references}

\begin{thebibliography}{33}%
\makeatletter
\providecommand \@ifxundefined [1]{%
 \@ifx{#1\undefined}
}%
\providecommand \@ifnum [1]{%
 \ifnum #1\expandafter \@firstoftwo
 \else \expandafter \@secondoftwo
 \fi
}%
\providecommand \@ifx [1]{%
 \ifx #1\expandafter \@firstoftwo
 \else \expandafter \@secondoftwo
 \fi
}%
\providecommand \natexlab [1]{#1}%
\providecommand \enquote  [1]{``#1''}%
\providecommand \bibnamefont  [1]{#1}%
\providecommand \bibfnamefont [1]{#1}%
\providecommand \citenamefont [1]{#1}%
\providecommand \href@noop [0]{\@secondoftwo}%
\providecommand \href [0]{\begingroup \@sanitize@url \@href}%
\providecommand \@href[1]{\@@startlink{#1}\@@href}%
\providecommand \@@href[1]{\endgroup#1\@@endlink}%
\providecommand \@sanitize@url [0]{\catcode `\\12\catcode `\$12\catcode
  `\&12\catcode `\#12\catcode `\^12\catcode `\_12\catcode `\%12\relax}%
\providecommand \@@startlink[1]{}%
\providecommand \@@endlink[0]{}%
\providecommand \url  [0]{\begingroup\@sanitize@url \@url }%
\providecommand \@url [1]{\endgroup\@href {#1}{\urlprefix }}%
\providecommand \urlprefix  [0]{URL }%
\providecommand \Eprint [0]{\href }%
\providecommand \doibase [0]{https://doi.org/}%
\providecommand \selectlanguage [0]{\@gobble}%
\providecommand \bibinfo  [0]{\@secondoftwo}%
\providecommand \bibfield  [0]{\@secondoftwo}%
\providecommand \translation [1]{[#1]}%
\providecommand \BibitemOpen [0]{}%
\providecommand \bibitemStop [0]{}%
\providecommand \bibitemNoStop [0]{.\EOS\space}%
\providecommand \EOS [0]{\spacefactor3000\relax}%
\providecommand \BibitemShut  [1]{\csname bibitem#1\endcsname}%
\let\auto@bib@innerbib\@empty
\bibitem [{\citenamefont {Caves}(1981)}]{Caves81}%
  \BibitemOpen
  \bibfield  {author} {\bibinfo {author} {\bibfnamefont {C.~M.}\ \bibnamefont
  {Caves}},\ }\bibfield  {title} {\bibinfo {title} {Quantum-mechanical noise in
  an interferometer},\ }\href {https://doi.org/10.1103/PhysRevD.23.1693}
  {\bibfield  {journal} {\bibinfo  {journal} {Phys. Rev. D}\ }\textbf {\bibinfo
  {volume} {23}},\ \bibinfo {pages} {1693} (\bibinfo {year}
  {1981})}\BibitemShut {NoStop}%
\bibitem [{\citenamefont {Caves}\ and\ \citenamefont
  {Schumaker}(1985)}]{Caves85P1}%
  \BibitemOpen
  \bibfield  {author} {\bibinfo {author} {\bibfnamefont {C.~M.}\ \bibnamefont
  {Caves}}\ and\ \bibinfo {author} {\bibfnamefont {B.~L.}\ \bibnamefont
  {Schumaker}},\ }\bibfield  {title} {\bibinfo {title} {New formalism for
  two-photon quantum optics. i. quadrature phases and squeezed states},\ }\href
  {https://doi.org/10.1103/PhysRevA.31.3068} {\bibfield  {journal} {\bibinfo
  {journal} {Phys. Rev. A}\ }\textbf {\bibinfo {volume} {31}},\ \bibinfo
  {pages} {3068} (\bibinfo {year} {1985})}\BibitemShut {NoStop}%
\bibitem [{\citenamefont {Schumaker}\ and\ \citenamefont
  {Caves}(1985)}]{Caves85P2}%
  \BibitemOpen
  \bibfield  {author} {\bibinfo {author} {\bibfnamefont {B.~L.}\ \bibnamefont
  {Schumaker}}\ and\ \bibinfo {author} {\bibfnamefont {C.~M.}\ \bibnamefont
  {Caves}},\ }\bibfield  {title} {\bibinfo {title} {New formalism for
  two-photon quantum optics. ii. mathematical foundation and compact
  notation},\ }\href {https://doi.org/10.1103/PhysRevA.31.3093} {\bibfield
  {journal} {\bibinfo  {journal} {Phys. Rev. A}\ }\textbf {\bibinfo {volume}
  {31}},\ \bibinfo {pages} {3093} (\bibinfo {year} {1985})}\BibitemShut
  {NoStop}%
\bibitem [{\citenamefont {Adhikari}(2014)}]{GravWavIntRev}%
  \BibitemOpen
  \bibfield  {author} {\bibinfo {author} {\bibfnamefont {R.~X.}\ \bibnamefont
  {Adhikari}},\ }\bibfield  {title} {\bibinfo {title} {Gravitational radiation
  detection with laser interferometry},\ }\href
  {https://doi.org/10.1103/RevModPhys.86.121} {\bibfield  {journal} {\bibinfo
  {journal} {Rev. Mod. Phys.}\ }\textbf {\bibinfo {volume} {86}},\ \bibinfo
  {pages} {121} (\bibinfo {year} {2014})}\BibitemShut {NoStop}%
\bibitem [{\citenamefont {Ma}\ \emph {et~al.}(2017)\citenamefont {Ma},
  \citenamefont {Miao}, \citenamefont {Pang}, \citenamefont {Evans},
  \citenamefont {Zhao}, \citenamefont {Harms}, \citenamefont {Schnabel},\ and\
  \citenamefont {Chen}}]{EntangPhaseEst}%
  \BibitemOpen
  \bibfield  {author} {\bibinfo {author} {\bibfnamefont {Y.}~\bibnamefont
  {Ma}}, \bibinfo {author} {\bibfnamefont {H.}~\bibnamefont {Miao}}, \bibinfo
  {author} {\bibfnamefont {B.~H.}\ \bibnamefont {Pang}}, \bibinfo {author}
  {\bibfnamefont {M.}~\bibnamefont {Evans}}, \bibinfo {author} {\bibfnamefont
  {C.}~\bibnamefont {Zhao}}, \bibinfo {author} {\bibfnamefont {J.}~\bibnamefont
  {Harms}}, \bibinfo {author} {\bibfnamefont {R.}~\bibnamefont {Schnabel}},\
  and\ \bibinfo {author} {\bibfnamefont {Y.}~\bibnamefont {Chen}},\ }\bibfield
  {title} {\bibinfo {title} {Proposal for gravitational-wave detection beyond
  the standard quantum limit through {EPR} entanglement},\ }\href
  {https://doi.org/10.1038/nphys4118} {\bibfield  {journal} {\bibinfo
  {journal} {Nat. Phys.}\ }\textbf {\bibinfo {volume} {13}},\ \bibinfo {pages}
  {776} (\bibinfo {year} {2017})}\BibitemShut {NoStop}%
\bibitem [{\citenamefont {Helstrom}(1969)}]{QFIhels}%
  \BibitemOpen
  \bibfield  {author} {\bibinfo {author} {\bibfnamefont {C.~W.}\ \bibnamefont
  {Helstrom}},\ }\bibfield  {title} {\bibinfo {title} {Quantum detection and
  estimation theory},\ }\href {https://doi.org/10.1007/BF01007479} {\bibfield
  {journal} {\bibinfo  {journal} {J. Stat. Phys.}\ }\textbf {\bibinfo {volume}
  {1}},\ \bibinfo {pages} {231} (\bibinfo {year} {1969})}\BibitemShut {NoStop}%
\bibitem [{\citenamefont {Paris}(2009)}]{QFIparis}%
  \BibitemOpen
  \bibfield  {author} {\bibinfo {author} {\bibfnamefont {M.~G.~A.}\
  \bibnamefont {Paris}},\ }\bibfield  {title} {\bibinfo {title} {Quantum
  estimation for quantum technology},\ }\href
  {https://doi.org/10.1142/S0219749909004839} {\bibfield  {journal} {\bibinfo
  {journal} {Int. J. Quantum Inf.}\ }\textbf {\bibinfo {volume} {7}},\ \bibinfo
  {pages} {125} (\bibinfo {year} {2009})}\BibitemShut {NoStop}%
\bibitem [{\citenamefont {Jarzyna}\ and\ \citenamefont
  {Demkowicz-Dobrza\ifmmode~\acute{n}\else \'{n}\fi{}ski}(2012)}]{QFIinterfer}%
  \BibitemOpen
  \bibfield  {author} {\bibinfo {author} {\bibfnamefont {M.}~\bibnamefont
  {Jarzyna}}\ and\ \bibinfo {author} {\bibfnamefont {R.}~\bibnamefont
  {Demkowicz-Dobrza\ifmmode~\acute{n}\else \'{n}\fi{}ski}},\ }\bibfield
  {title} {\bibinfo {title} {Quantum interferometry with and without an
  external phase reference},\ }\href
  {https://doi.org/10.1103/PhysRevA.85.011801} {\bibfield  {journal} {\bibinfo
  {journal} {Phys. Rev. A}\ }\textbf {\bibinfo {volume} {85}},\ \bibinfo
  {pages} {011801} (\bibinfo {year} {2012})}\BibitemShut {NoStop}%
\bibitem [{\citenamefont {Horoshko}\ and\ \citenamefont
  {Jelezko}(2026)}]{QFIintOneMode26}%
  \BibitemOpen
  \bibfield  {author} {\bibinfo {author} {\bibfnamefont {D.~B.}\ \bibnamefont
  {Horoshko}}\ and\ \bibinfo {author} {\bibfnamefont {F.}~\bibnamefont
  {Jelezko}},\ }\bibfield  {title} {\bibinfo {title} {Quantum limit of
  precision for phase estimation in squeezing-enhanced interferometry with a
  single-mode readout},\ }\href@noop {} {\bibfield  {journal} {\bibinfo
  {journal} {arXiv preprint arXiv:2603.07556}\ } (\bibinfo {year}
  {2026})}\BibitemShut {NoStop}%
\bibitem [{\citenamefont {Pezz\'e}\ and\ \citenamefont
  {Smerzi}(2008)}]{IntOptMeas}%
  \BibitemOpen
  \bibfield  {author} {\bibinfo {author} {\bibfnamefont {L.}~\bibnamefont
  {Pezz\'e}}\ and\ \bibinfo {author} {\bibfnamefont {A.}~\bibnamefont
  {Smerzi}},\ }\bibfield  {title} {\bibinfo {title} {Mach-zehnder
  interferometry at the heisenberg limit with coherent and squeezed-vacuum
  light},\ }\href {https://doi.org/10.1103/PhysRevLett.100.073601} {\bibfield
  {journal} {\bibinfo  {journal} {Phys. Rev. Lett.}\ }\textbf {\bibinfo
  {volume} {100}},\ \bibinfo {pages} {073601} (\bibinfo {year}
  {2008})}\BibitemShut {NoStop}%
\bibitem [{\citenamefont {Lang}\ and\ \citenamefont
  {Caves}(2013)}]{IntOptState}%
  \BibitemOpen
  \bibfield  {author} {\bibinfo {author} {\bibfnamefont {M.~D.}\ \bibnamefont
  {Lang}}\ and\ \bibinfo {author} {\bibfnamefont {C.~M.}\ \bibnamefont
  {Caves}},\ }\bibfield  {title} {\bibinfo {title} {Optimal quantum-enhanced
  interferometry using a laser power source},\ }\href
  {https://doi.org/10.1103/PhysRevLett.111.173601} {\bibfield  {journal}
  {\bibinfo  {journal} {Phys. Rev. Lett.}\ }\textbf {\bibinfo {volume} {111}},\
  \bibinfo {pages} {173601} (\bibinfo {year} {2013})}\BibitemShut {NoStop}%
\bibitem [{\citenamefont {Amelino-Camelia}(1999)}]{amelino99}%
  \BibitemOpen
  \bibfield  {author} {\bibinfo {author} {\bibfnamefont {G.}~\bibnamefont
  {Amelino-Camelia}},\ }\bibfield  {title} {\bibinfo {title} {Gravity-wave
  interferometers as quantum-gravity detectors},\ }\href
  {https://doi.org/10.1038/18377} {\bibfield  {journal} {\bibinfo  {journal}
  {Nature}\ }\textbf {\bibinfo {volume} {398}},\ \bibinfo {pages} {216}
  (\bibinfo {year} {1999})}\BibitemShut {NoStop}%
\bibitem [{\citenamefont {Amelino-Camelia}(2000)}]{amelinoPRD}%
  \BibitemOpen
  \bibfield  {author} {\bibinfo {author} {\bibfnamefont {G.}~\bibnamefont
  {Amelino-Camelia}},\ }\bibfield  {title} {\bibinfo {title} {Gravity-wave
  interferometers as probes of a low-energy effective quantum gravity},\ }\href
  {https://doi.org/10.1103/PhysRevD.62.024015} {\bibfield  {journal} {\bibinfo
  {journal} {Phys. Rev. D}\ }\textbf {\bibinfo {volume} {62}},\ \bibinfo
  {pages} {024015} (\bibinfo {year} {2000})}\BibitemShut {NoStop}%
\bibitem [{\citenamefont {Chou}\ \emph {et~al.}(2017)\citenamefont {Chou},
  \citenamefont {Glass}, \citenamefont {Gustafson}, \citenamefont {Hogan},
  \citenamefont {Kamai}, \citenamefont {Kwon}, \citenamefont {Lanza},
  \citenamefont {McCuller}, \citenamefont {Meyer}, \citenamefont {Richardson},
  \citenamefont {Stoughton}, \citenamefont {Tomlin},\ and\ \citenamefont
  {Weiss}}]{HoloData}%
  \BibitemOpen
  \bibfield  {author} {\bibinfo {author} {\bibfnamefont {A.}~\bibnamefont
  {Chou}}, \bibinfo {author} {\bibfnamefont {H.}~\bibnamefont {Glass}},
  \bibinfo {author} {\bibfnamefont {H.~R.}\ \bibnamefont {Gustafson}}, \bibinfo
  {author} {\bibfnamefont {C.~J.}\ \bibnamefont {Hogan}}, \bibinfo {author}
  {\bibfnamefont {B.~L.}\ \bibnamefont {Kamai}}, \bibinfo {author}
  {\bibfnamefont {O.}~\bibnamefont {Kwon}}, \bibinfo {author} {\bibfnamefont
  {R.}~\bibnamefont {Lanza}}, \bibinfo {author} {\bibfnamefont
  {L.}~\bibnamefont {McCuller}}, \bibinfo {author} {\bibfnamefont {S.~S.}\
  \bibnamefont {Meyer}}, \bibinfo {author} {\bibfnamefont {J.~W.}\ \bibnamefont
  {Richardson}}, \bibinfo {author} {\bibfnamefont {C.}~\bibnamefont
  {Stoughton}}, \bibinfo {author} {\bibfnamefont {R.}~\bibnamefont {Tomlin}},\
  and\ \bibinfo {author} {\bibfnamefont {R.}~\bibnamefont {Weiss}} (\bibinfo
  {collaboration} {Holometer Collaboration}),\ }\bibfield  {title} {\bibinfo
  {title} {Interferometric constraints on quantum geometrical shear noise
  correlations},\ }\href {https://doi.org/10.1088/1361-6382/aa7bd3} {\bibfield
  {journal} {\bibinfo  {journal} {Classical Quant. Grav.}\ }\textbf {\bibinfo
  {volume} {34}},\ \bibinfo {pages} {165005} (\bibinfo {year}
  {2017})}\BibitemShut {NoStop}%
\bibitem [{\citenamefont {Richardson}\ \emph {et~al.}(2021)\citenamefont
  {Richardson}, \citenamefont {Kwon}, \citenamefont {Gustafson}, \citenamefont
  {Hogan}, \citenamefont {Kamai}, \citenamefont {McCuller}, \citenamefont
  {Meyer}, \citenamefont {Stoughton}, \citenamefont {Tomlin},\ and\
  \citenamefont {Weiss}}]{bentHolo}%
  \BibitemOpen
  \bibfield  {author} {\bibinfo {author} {\bibfnamefont {J.~W.}\ \bibnamefont
  {Richardson}}, \bibinfo {author} {\bibfnamefont {O.}~\bibnamefont {Kwon}},
  \bibinfo {author} {\bibfnamefont {H.~R.}\ \bibnamefont {Gustafson}}, \bibinfo
  {author} {\bibfnamefont {C.}~\bibnamefont {Hogan}}, \bibinfo {author}
  {\bibfnamefont {B.~L.}\ \bibnamefont {Kamai}}, \bibinfo {author}
  {\bibfnamefont {L.~P.}\ \bibnamefont {McCuller}}, \bibinfo {author}
  {\bibfnamefont {S.~S.}\ \bibnamefont {Meyer}}, \bibinfo {author}
  {\bibfnamefont {C.}~\bibnamefont {Stoughton}}, \bibinfo {author}
  {\bibfnamefont {R.~E.}\ \bibnamefont {Tomlin}},\ and\ \bibinfo {author}
  {\bibfnamefont {R.}~\bibnamefont {Weiss}},\ }\bibfield  {title} {\bibinfo
  {title} {Interferometric constraints on spacelike coherent rotational
  fluctuations},\ }\href {https://doi.org/10.1103/PhysRevLett.126.241301}
  {\bibfield  {journal} {\bibinfo  {journal} {Phys. Rev. Lett.}\ }\textbf
  {\bibinfo {volume} {126}},\ \bibinfo {pages} {241301} (\bibinfo {year}
  {2021})}\BibitemShut {NoStop}%
\bibitem [{\citenamefont {Patra}\ \emph {et~al.}(2025)\citenamefont {Patra},
  \citenamefont {Aiello}, \citenamefont {Ejlli}, \citenamefont {Griffiths},
  \citenamefont {James}, \citenamefont {Kuntimaddi}, \citenamefont {Kwon},
  \citenamefont {Schwartz}, \citenamefont {Vahlbruch}, \citenamefont
  {Vermeulen}, \citenamefont {Kokeyama}, \citenamefont {Dooley},\ and\
  \citenamefont {Grote}}]{QUESTdata}%
  \BibitemOpen
  \bibfield  {author} {\bibinfo {author} {\bibfnamefont {A.}~\bibnamefont
  {Patra}}, \bibinfo {author} {\bibfnamefont {L.}~\bibnamefont {Aiello}},
  \bibinfo {author} {\bibfnamefont {A.}~\bibnamefont {Ejlli}}, \bibinfo
  {author} {\bibfnamefont {W.~L.}\ \bibnamefont {Griffiths}}, \bibinfo {author}
  {\bibfnamefont {A.~L.}\ \bibnamefont {James}}, \bibinfo {author}
  {\bibfnamefont {N.}~\bibnamefont {Kuntimaddi}}, \bibinfo {author}
  {\bibfnamefont {O.}~\bibnamefont {Kwon}}, \bibinfo {author} {\bibfnamefont
  {E.}~\bibnamefont {Schwartz}}, \bibinfo {author} {\bibfnamefont
  {H.}~\bibnamefont {Vahlbruch}}, \bibinfo {author} {\bibfnamefont {S.~M.}\
  \bibnamefont {Vermeulen}}, \bibinfo {author} {\bibfnamefont {K.}~\bibnamefont
  {Kokeyama}}, \bibinfo {author} {\bibfnamefont {K.~L.}\ \bibnamefont
  {Dooley}},\ and\ \bibinfo {author} {\bibfnamefont {H.}~\bibnamefont
  {Grote}},\ }\bibfield  {title} {\bibinfo {title} {Broadband limits on
  stochastic length fluctuations from a pair of table-top interferometers},\
  }\href {https://doi.org/10.1103/61j9-cjkk} {\bibfield  {journal} {\bibinfo
  {journal} {Phys. Rev. Lett.}\ }\textbf {\bibinfo {volume} {135}},\ \bibinfo
  {pages} {101402} (\bibinfo {year} {2025})}\BibitemShut {NoStop}%
\bibitem [{\citenamefont {Ruo~Berchera}\ \emph {et~al.}(2013)\citenamefont
  {Ruo~Berchera}, \citenamefont {Degiovanni}, \citenamefont {Olivares},\ and\
  \citenamefont {Genovese}}]{QLtIRB1}%
  \BibitemOpen
  \bibfield  {author} {\bibinfo {author} {\bibfnamefont {I.}~\bibnamefont
  {Ruo~Berchera}}, \bibinfo {author} {\bibfnamefont {I.~P.}\ \bibnamefont
  {Degiovanni}}, \bibinfo {author} {\bibfnamefont {S.}~\bibnamefont
  {Olivares}},\ and\ \bibinfo {author} {\bibfnamefont {M.}~\bibnamefont
  {Genovese}},\ }\bibfield  {title} {\bibinfo {title} {Quantum light in coupled
  interferometers for quantum gravity tests},\ }\href
  {https://doi.org/10.1103/PhysRevLett.110.213601} {\bibfield  {journal}
  {\bibinfo  {journal} {Phys. Rev. Lett.}\ }\textbf {\bibinfo {volume} {110}},\
  \bibinfo {pages} {213601} (\bibinfo {year} {2013})}\BibitemShut {NoStop}%
\bibitem [{\citenamefont {Ruo-Berchera}\ \emph {et~al.}(2015)\citenamefont
  {Ruo-Berchera}, \citenamefont {Degiovanni}, \citenamefont {Olivares},
  \citenamefont {Samantaray}, \citenamefont {Traina},\ and\ \citenamefont
  {Genovese}}]{QLtIRB2}%
  \BibitemOpen
  \bibfield  {author} {\bibinfo {author} {\bibfnamefont {I.}~\bibnamefont
  {Ruo-Berchera}}, \bibinfo {author} {\bibfnamefont {I.~P.}\ \bibnamefont
  {Degiovanni}}, \bibinfo {author} {\bibfnamefont {S.}~\bibnamefont
  {Olivares}}, \bibinfo {author} {\bibfnamefont {N.}~\bibnamefont
  {Samantaray}}, \bibinfo {author} {\bibfnamefont {P.}~\bibnamefont {Traina}},\
  and\ \bibinfo {author} {\bibfnamefont {M.}~\bibnamefont {Genovese}},\
  }\bibfield  {title} {\bibinfo {title} {One- and two-mode squeezed light in
  correlated interferometry},\ }\href
  {https://doi.org/10.1103/PhysRevA.92.053821} {\bibfield  {journal} {\bibinfo
  {journal} {Phys. Rev. A}\ }\textbf {\bibinfo {volume} {92}},\ \bibinfo
  {pages} {053821} (\bibinfo {year} {2015})}\BibitemShut {NoStop}%
\bibitem [{\citenamefont {Gardner}\ \emph {et~al.}(2025)\citenamefont
  {Gardner}, \citenamefont {Gefen}, \citenamefont {Haine}, \citenamefont
  {Hope}, \citenamefont {Preskill}, \citenamefont {Chen},\ and\ \citenamefont
  {McCuller}}]{gardner2025}%
  \BibitemOpen
  \bibfield  {author} {\bibinfo {author} {\bibfnamefont {J.~W.}\ \bibnamefont
  {Gardner}}, \bibinfo {author} {\bibfnamefont {T.}~\bibnamefont {Gefen}},
  \bibinfo {author} {\bibfnamefont {S.~A.}\ \bibnamefont {Haine}}, \bibinfo
  {author} {\bibfnamefont {J.~J.}\ \bibnamefont {Hope}}, \bibinfo {author}
  {\bibfnamefont {J.}~\bibnamefont {Preskill}}, \bibinfo {author}
  {\bibfnamefont {Y.}~\bibnamefont {Chen}},\ and\ \bibinfo {author}
  {\bibfnamefont {L.}~\bibnamefont {McCuller}},\ }\bibfield  {title} {\bibinfo
  {title} {Stochastic waveform estimation at the fundamental quantum limit},\
  }\href {https://doi.org/10.1103/h91r-4ws9} {\bibfield  {journal} {\bibinfo
  {journal} {PRX Quantum}\ }\textbf {\bibinfo {volume} {6}},\ \bibinfo {pages}
  {030311} (\bibinfo {year} {2025})}\BibitemShut {NoStop}%
\bibitem [{\citenamefont {Yurke}\ \emph {et~al.}(1986)\citenamefont {Yurke},
  \citenamefont {McCall},\ and\ \citenamefont {Klauder}}]{SU11YMK}%
  \BibitemOpen
  \bibfield  {author} {\bibinfo {author} {\bibfnamefont {B.}~\bibnamefont
  {Yurke}}, \bibinfo {author} {\bibfnamefont {S.~L.}\ \bibnamefont {McCall}},\
  and\ \bibinfo {author} {\bibfnamefont {J.~R.}\ \bibnamefont {Klauder}},\
  }\bibfield  {title} {\bibinfo {title} {{SU(2)} and {SU(1,1)}
  interferometers},\ }\href {https://doi.org/10.1103/PhysRevA.33.4033}
  {\bibfield  {journal} {\bibinfo  {journal} {Phys. Rev. A}\ }\textbf {\bibinfo
  {volume} {33}},\ \bibinfo {pages} {4033} (\bibinfo {year}
  {1986})}\BibitemShut {NoStop}%
\bibitem [{\citenamefont {Caves}(2020)}]{SU11Caves}%
  \BibitemOpen
  \bibfield  {author} {\bibinfo {author} {\bibfnamefont {C.~M.}\ \bibnamefont
  {Caves}},\ }\bibfield  {title} {\bibinfo {title} {Reframing {SU(1,1)}
  interferometry},\ }\href
  {https://doi.org/https://doi.org/10.1002/qute.201900138} {\bibfield
  {journal} {\bibinfo  {journal} {Adv. Quantum Technol.}\ }\textbf {\bibinfo
  {volume} {3}},\ \bibinfo {pages} {1900138} (\bibinfo {year}
  {2020})}\BibitemShut {NoStop}%
\bibitem [{\citenamefont {Jing}\ \emph {et~al.}(2011)\citenamefont {Jing},
  \citenamefont {Liu}, \citenamefont {Zhou}, \citenamefont {Ou},\ and\
  \citenamefont {Zhang}}]{SU11Expt1}%
  \BibitemOpen
  \bibfield  {author} {\bibinfo {author} {\bibfnamefont {J.}~\bibnamefont
  {Jing}}, \bibinfo {author} {\bibfnamefont {C.}~\bibnamefont {Liu}}, \bibinfo
  {author} {\bibfnamefont {Z.}~\bibnamefont {Zhou}}, \bibinfo {author}
  {\bibfnamefont {Z.~Y.}\ \bibnamefont {Ou}},\ and\ \bibinfo {author}
  {\bibfnamefont {W.}~\bibnamefont {Zhang}},\ }\bibfield  {title} {\bibinfo
  {title} {Realization of a nonlinear interferometer with parametric
  amplifiers},\ }\href {https://doi.org/10.1063/1.3606549} {\bibfield
  {journal} {\bibinfo  {journal} {Appl. Phys. Lett.}\ }\textbf {\bibinfo
  {volume} {99}},\ \bibinfo {pages} {011110} (\bibinfo {year}
  {2011})}\BibitemShut {NoStop}%
\bibitem [{\citenamefont {Hudelist}\ \emph {et~al.}(2014)\citenamefont
  {Hudelist}, \citenamefont {Kong}, \citenamefont {Liu}, \citenamefont {Jing},
  \citenamefont {Ou},\ and\ \citenamefont {Zhang}}]{SU11Expt2}%
  \BibitemOpen
  \bibfield  {author} {\bibinfo {author} {\bibfnamefont {F.}~\bibnamefont
  {Hudelist}}, \bibinfo {author} {\bibfnamefont {J.}~\bibnamefont {Kong}},
  \bibinfo {author} {\bibfnamefont {C.}~\bibnamefont {Liu}}, \bibinfo {author}
  {\bibfnamefont {J.}~\bibnamefont {Jing}}, \bibinfo {author} {\bibfnamefont
  {Z.}~\bibnamefont {Ou}},\ and\ \bibinfo {author} {\bibfnamefont
  {W.}~\bibnamefont {Zhang}},\ }\bibfield  {title} {\bibinfo {title} {Quantum
  metrology with parametric amplifier-based photon correlation
  interferometers},\ }\href {https://doi.org/10.1038/ncomms4049} {\bibfield
  {journal} {\bibinfo  {journal} {Nat. Commun.}\ }\textbf {\bibinfo {volume}
  {5}},\ \bibinfo {pages} {3049} (\bibinfo {year} {2014})}\BibitemShut
  {NoStop}%
\bibitem [{\citenamefont {Machado}\ \emph {et~al.}(2020)\citenamefont
  {Machado}, \citenamefont {Frascella}, \citenamefont {Torres},\ and\
  \citenamefont {Chekhova}}]{SU11Expt3}%
  \BibitemOpen
  \bibfield  {author} {\bibinfo {author} {\bibfnamefont {G.~J.}\ \bibnamefont
  {Machado}}, \bibinfo {author} {\bibfnamefont {G.}~\bibnamefont {Frascella}},
  \bibinfo {author} {\bibfnamefont {J.~P.}\ \bibnamefont {Torres}},\ and\
  \bibinfo {author} {\bibfnamefont {M.~V.}\ \bibnamefont {Chekhova}},\
  }\bibfield  {title} {\bibinfo {title} {Optical coherence tomography with a
  nonlinear interferometer in the high parametric gain regime},\ }\href
  {https://doi.org/10.1063/5.0016259} {\bibfield  {journal} {\bibinfo
  {journal} {Appl. Phys. Lett.}\ }\textbf {\bibinfo {volume} {117}},\ \bibinfo
  {pages} {094002} (\bibinfo {year} {2020})}\BibitemShut {NoStop}%
\bibitem [{\citenamefont {Manceau}\ \emph {et~al.}(2017)\citenamefont
  {Manceau}, \citenamefont {Leuchs}, \citenamefont {Khalili},\ and\
  \citenamefont {Chekhova}}]{SU11PRL}%
  \BibitemOpen
  \bibfield  {author} {\bibinfo {author} {\bibfnamefont {M.}~\bibnamefont
  {Manceau}}, \bibinfo {author} {\bibfnamefont {G.}~\bibnamefont {Leuchs}},
  \bibinfo {author} {\bibfnamefont {F.}~\bibnamefont {Khalili}},\ and\ \bibinfo
  {author} {\bibfnamefont {M.}~\bibnamefont {Chekhova}},\ }\bibfield  {title}
  {\bibinfo {title} {Detection loss tolerant supersensitive phase measurement
  with an {SU(1,1)} interferometer},\ }\href
  {https://doi.org/10.1103/PhysRevLett.119.223604} {\bibfield  {journal}
  {\bibinfo  {journal} {Phys. Rev. Lett.}\ }\textbf {\bibinfo {volume} {119}},\
  \bibinfo {pages} {223604} (\bibinfo {year} {2017})}\BibitemShut {NoStop}%
\bibitem [{\citenamefont {Zheng}\ \emph {et~al.}(2020)\citenamefont {Zheng},
  \citenamefont {Mi}, \citenamefont {Wang}, \citenamefont {Xu}, \citenamefont
  {Hu}, \citenamefont {Liu}, \citenamefont {Lou}, \citenamefont {Jing},\ and\
  \citenamefont {Zhang}}]{SU11StochPhase}%
  \BibitemOpen
  \bibfield  {author} {\bibinfo {author} {\bibfnamefont {K.}~\bibnamefont
  {Zheng}}, \bibinfo {author} {\bibfnamefont {M.}~\bibnamefont {Mi}}, \bibinfo
  {author} {\bibfnamefont {B.}~\bibnamefont {Wang}}, \bibinfo {author}
  {\bibfnamefont {L.}~\bibnamefont {Xu}}, \bibinfo {author} {\bibfnamefont
  {L.}~\bibnamefont {Hu}}, \bibinfo {author} {\bibfnamefont {S.}~\bibnamefont
  {Liu}}, \bibinfo {author} {\bibfnamefont {Y.}~\bibnamefont {Lou}}, \bibinfo
  {author} {\bibfnamefont {J.}~\bibnamefont {Jing}},\ and\ \bibinfo {author}
  {\bibfnamefont {L.}~\bibnamefont {Zhang}},\ }\bibfield  {title} {\bibinfo
  {title} {Quantum-enhanced stochastic phase estimation with the {SU(1,1)}
  interferometer},\ }\href {https://doi.org/10.1364/PRJ.395682} {\bibfield
  {journal} {\bibinfo  {journal} {Photon. Res.}\ }\textbf {\bibinfo {volume}
  {8}},\ \bibinfo {pages} {1653} (\bibinfo {year} {2020})}\BibitemShut
  {NoStop}%
\bibitem [{\citenamefont {Korobko}\ \emph {et~al.}(2017)\citenamefont
  {Korobko}, \citenamefont {Kleybolte}, \citenamefont {Ast}, \citenamefont
  {Miao}, \citenamefont {Chen},\ and\ \citenamefont {Schnabel}}]{IntSq1}%
  \BibitemOpen
  \bibfield  {author} {\bibinfo {author} {\bibfnamefont {M.}~\bibnamefont
  {Korobko}}, \bibinfo {author} {\bibfnamefont {L.}~\bibnamefont {Kleybolte}},
  \bibinfo {author} {\bibfnamefont {S.}~\bibnamefont {Ast}}, \bibinfo {author}
  {\bibfnamefont {H.}~\bibnamefont {Miao}}, \bibinfo {author} {\bibfnamefont
  {Y.}~\bibnamefont {Chen}},\ and\ \bibinfo {author} {\bibfnamefont
  {R.}~\bibnamefont {Schnabel}},\ }\bibfield  {title} {\bibinfo {title}
  {Beating the standard sensitivity-bandwidth limit of cavity-enhanced
  interferometers with internal squeezed-light generation},\ }\href
  {https://doi.org/10.1103/PhysRevLett.118.143601} {\bibfield  {journal}
  {\bibinfo  {journal} {Phys. Rev. Lett.}\ }\textbf {\bibinfo {volume} {118}},\
  \bibinfo {pages} {143601} (\bibinfo {year} {2017})}\BibitemShut {NoStop}%
\bibitem [{\citenamefont {Gardner}\ \emph {et~al.}(2022)\citenamefont
  {Gardner}, \citenamefont {Yap}, \citenamefont {Adya}, \citenamefont {Chua},
  \citenamefont {Slagmolen},\ and\ \citenamefont {McClelland}}]{IntSq2}%
  \BibitemOpen
  \bibfield  {author} {\bibinfo {author} {\bibfnamefont {J.~W.}\ \bibnamefont
  {Gardner}}, \bibinfo {author} {\bibfnamefont {M.~J.}\ \bibnamefont {Yap}},
  \bibinfo {author} {\bibfnamefont {V.}~\bibnamefont {Adya}}, \bibinfo {author}
  {\bibfnamefont {S.}~\bibnamefont {Chua}}, \bibinfo {author} {\bibfnamefont
  {B.~J.~J.}\ \bibnamefont {Slagmolen}},\ and\ \bibinfo {author} {\bibfnamefont
  {D.~E.}\ \bibnamefont {McClelland}},\ }\bibfield  {title} {\bibinfo {title}
  {Nondegenerate internal squeezing: An all-optical, loss-resistant quantum
  technique for gravitational-wave detection},\ }\href
  {https://doi.org/10.1103/PhysRevD.106.L041101} {\bibfield  {journal}
  {\bibinfo  {journal} {Phys. Rev. D}\ }\textbf {\bibinfo {volume} {106}},\
  \bibinfo {pages} {L041101} (\bibinfo {year} {2022})}\BibitemShut {NoStop}%
\bibitem [{\citenamefont {Vermeulen}\ \emph {et~al.}(2026)\citenamefont
  {Vermeulen}, \citenamefont {Koca},\ and\ \citenamefont {McCuller}}]{IntSq3}%
  \BibitemOpen
  \bibfield  {author} {\bibinfo {author} {\bibfnamefont {S.~M.}\ \bibnamefont
  {Vermeulen}}, \bibinfo {author} {\bibfnamefont {U.~S.}\ \bibnamefont
  {Koca}},\ and\ \bibinfo {author} {\bibfnamefont {L.}~\bibnamefont
  {McCuller}},\ }\bibfield  {title} {\bibinfo {title} {Bidirectional internal
  squeezing for gravitational-wave detectors},\ }\href
  {https://doi.org/10.48550/arXiv.2605.16512} {\bibfield  {journal} {\bibinfo
  {journal} {arXiv preprint arXiv:2605.16512}\ } (\bibinfo {year}
  {2026})}\BibitemShut {NoStop}%
\bibitem [{\citenamefont {Monras}(2013)}]{QFI}%
  \BibitemOpen
  \bibfield  {author} {\bibinfo {author} {\bibfnamefont {A.}~\bibnamefont
  {Monras}},\ }\bibfield  {title} {\bibinfo {title} {Phase space formalism for
  quantum estimation of gaussian states},\ }\href
  {https://doi.org/10.48550/arXiv.1303.3682} {\bibfield  {journal} {\bibinfo
  {journal} {arXiv preprint arXiv:1303.3682}\ } (\bibinfo {year}
  {2013})}\BibitemShut {NoStop}%
\bibitem [{\citenamefont {Sharmila}\ \emph {et~al.}(2026)\citenamefont
  {Sharmila}, \citenamefont {Vermeulen},\ and\ \citenamefont {Datta}}]{shar25}%
  \BibitemOpen
  \bibfield  {author} {\bibinfo {author} {\bibfnamefont {B.}~\bibnamefont
  {Sharmila}}, \bibinfo {author} {\bibfnamefont {S.~M.}\ \bibnamefont
  {Vermeulen}},\ and\ \bibinfo {author} {\bibfnamefont {A.}~\bibnamefont
  {Datta}},\ }\bibfield  {title} {\bibinfo {title} {Signatures of correlation
  of spacetime fluctuations in laser interferometers},\ }\href
  {https://doi.org/10.1038/s41467-025-67313-3} {\bibfield  {journal} {\bibinfo
  {journal} {Nat. Commun.}\ }\textbf {\bibinfo {volume} {17}},\ \bibinfo
  {pages} {701} (\bibinfo {year} {2026})}\BibitemShut {NoStop}%
\bibitem [{\citenamefont {Vermeulen}\ \emph {et~al.}(2025)\citenamefont
  {Vermeulen}, \citenamefont {Cullen}, \citenamefont {Grass}, \citenamefont
  {MacMillan}, \citenamefont {Ramirez}, \citenamefont {Wack}, \citenamefont
  {Korzh}, \citenamefont {Lee}, \citenamefont {Zurek}, \citenamefont
  {Stoughton},\ and\ \citenamefont {McCuller}}]{smv24}%
  \BibitemOpen
  \bibfield  {author} {\bibinfo {author} {\bibfnamefont {S.~M.}\ \bibnamefont
  {Vermeulen}}, \bibinfo {author} {\bibfnamefont {T.}~\bibnamefont {Cullen}},
  \bibinfo {author} {\bibfnamefont {D.}~\bibnamefont {Grass}}, \bibinfo
  {author} {\bibfnamefont {I.~A.~O.}\ \bibnamefont {MacMillan}}, \bibinfo
  {author} {\bibfnamefont {A.~J.}\ \bibnamefont {Ramirez}}, \bibinfo {author}
  {\bibfnamefont {J.}~\bibnamefont {Wack}}, \bibinfo {author} {\bibfnamefont
  {B.}~\bibnamefont {Korzh}}, \bibinfo {author} {\bibfnamefont {V.~S.~H.}\
  \bibnamefont {Lee}}, \bibinfo {author} {\bibfnamefont {K.~M.}\ \bibnamefont
  {Zurek}}, \bibinfo {author} {\bibfnamefont {C.}~\bibnamefont {Stoughton}},\
  and\ \bibinfo {author} {\bibfnamefont {L.}~\bibnamefont {McCuller}},\
  }\bibfield  {title} {\bibinfo {title} {Photon-counting interferometry to
  detect geontropic space-time fluctuations with {GQuEST}},\ }\href
  {https://doi.org/10.1103/PhysRevX.15.011034} {\bibfield  {journal} {\bibinfo
  {journal} {Phys. Rev. X}\ }\textbf {\bibinfo {volume} {15}},\ \bibinfo
  {pages} {011034} (\bibinfo {year} {2025})}\BibitemShut {NoStop}%
\bibitem [{\citenamefont {Hong}\ \emph {et~al.}(2025)\citenamefont {Hong},
  \citenamefont {Feldman}, \citenamefont {Marvinney}, \citenamefont {Lee},
  \citenamefont {Lee}, \citenamefont {Febbraro}, \citenamefont {Marino},\ and\
  \citenamefont {Pooser}}]{SU11Expt4}%
  \BibitemOpen
  \bibfield  {author} {\bibinfo {author} {\bibfnamefont {S.}~\bibnamefont
  {Hong}}, \bibinfo {author} {\bibfnamefont {M.~A.}\ \bibnamefont {Feldman}},
  \bibinfo {author} {\bibfnamefont {C.~E.}\ \bibnamefont {Marvinney}}, \bibinfo
  {author} {\bibfnamefont {D.}~\bibnamefont {Lee}}, \bibinfo {author}
  {\bibfnamefont {C.}~\bibnamefont {Lee}}, \bibinfo {author} {\bibfnamefont
  {M.~T.}\ \bibnamefont {Febbraro}}, \bibinfo {author} {\bibfnamefont {A.~M.}\
  \bibnamefont {Marino}},\ and\ \bibinfo {author} {\bibfnamefont {R.~C.}\
  \bibnamefont {Pooser}},\ }\bibfield  {title} {\bibinfo {title}
  {Quantum-enhanced distributed phase sensing with a truncated {SU(1,1)}
  interferometer},\ }\href {https://doi.org/10.1103/PhysRevResearch.7.023231}
  {\bibfield  {journal} {\bibinfo  {journal} {Phys. Rev. Res.}\ }\textbf
  {\bibinfo {volume} {7}},\ \bibinfo {pages} {023231} (\bibinfo {year}
  {2025})}\BibitemShut {NoStop}%
\end{thebibliography}%

\end{document}